\documentclass[%
    reprint,
    superscriptaddress,
    longbibliography,
    nofootinbib,
    amsmath,
    amssymb,
    aps,
    pra,
    floatfix,
]{revtex4-2}

\usepackage{graphicx}
\usepackage{dcolumn}
\usepackage{bm}
\usepackage[utf8]{inputenc}
\usepackage{acronym}
\usepackage{graphicx}
\usepackage{amsmath,amsfonts,amssymb}
\usepackage{braket}
\usepackage{tikz}
\usetikzlibrary{quantikz}
\usepackage{acronym}
\usepackage{xspace}
\usepackage{siunitx}
\usepackage{booktabs}
\usepackage[hidelinks]{hyperref}
\usepackage{nicefrac}       
\usepackage{microtype}      
\usepackage{mathtools}
\usepackage{listings}
\usepackage{siunitx}
\usepackage{multirow}

\begin{document}

\title{Advantages of density in tensor network geometries for gradient based training}
\author{Sergi Masot-Llima}
\affiliation{Barcelona Supercomputing Center, 08034 Barcelona, Spain}
\author{Artur Garcia-Saez}
\thanks{Contact author: \href{mailto:artur.garcia@bsc.es}{artur.garcia@bsc.es}}
\affiliation{Barcelona Supercomputing Center, 08034 Barcelona, Spain}
\affiliation{Qilimanjaro Quantum Tech, 08019 Barcelona, Spain}

\date{\today}

\begin{abstract}
Tensor networks are a very powerful data structure tool originating from quantum system simulations. In recent years, they have seen increased use in machine learning, mostly in trainings with gradient-based techniques, due to their flexibility and performance exploiting hardware acceleration. As ansätze, tensor networks can be used with flexible geometries, and it is known that for highly regular ones their dimensionality has a large impact in performance and representation power. For heterogeneous structures, however, these effects are not completely characterized. In this article, we train tensor networks with different geometries to encode a random quantum state, and see that densely connected structures achieve better infidelities than more sparse structures, with higher success rates and less time. Additionally, we give some general insight on how to improve memory requirements on these sparse structures and its impact on the trainings. Finally, as we use HPC resources for the calculations, we discuss the requirements for this approach and showcase performance improvements with GPU acceleration on a last-generation supercomputer.

\end{abstract}
\maketitle

\section{Introduction}


Tensor networks (TN) provide a characterization of quantum states with a connection to physical states of local Hamiltonians\cite{ciracMatrixProductStates2021} that may be efficiently represented classically. They have become an indispensable tool in the field of quantum simulation, as they offer highly developed methods that are parallelizable for large scale computations. For example, they can be used directly for circuit simulation \cite{Markov_2008}, both with large network contraction heuristics \cite{Gray_2024} or with statevector with bounded bond dimension \cite{Vidal_2003}, as well as for condensed matter computations with 1D \cite{mccullochInfiniteSizeDensity2008d} and higher dimensional methods \cite{vidalEntanglementRenormalization2007,verstraete2004renormalizationalgorithmsquantummanybody}. Recently, they have also played a role in quantum advantage proofs against cutting-edge Noisy Intermediate-Scale Quantum devices \cite{Begu_i__2024}. 

The particular flavour of 1D TN, namely Matrix Product States (MPS)\cite{Verstraete_2008}, has been identified as the variational structure operated by the celebrated DMRG algorithm \cite{Schollw_ck_2011}. Moreover, their flexibility in storing data and the usefulness of the bond dimension $\chi$, a parameter that controls how much correlation is allowed, has brought this tool to see success in machine learning as well \cite{stoudenmire2017,TNAD_paper} using gradient-based tools (either first or second order), similarly to classical machine learning.

The choice of the TN geometry in all these applications is related to how different systems present different entanglement structures: networks with a large complexity might be able to represent desirable states better. On the other hand, the use of highly intricate TN structures is limited by the computational cost of contractions and increased memory requirements. The interplay between these behaviours is crucial when choosing a strategy for solving a problem, yet the impact on the training performance is not completely characterised despite being relevant for contraction strategies \cite{Gray_2024}, numerical implementations \cite{evenbly2022practicalguidenumericalimplementation}, large scale computations \cite{PhysRevA.104.032603} and the appearance of barren plateaus \cite{Cervero_Mart_n_2023,PhysRevLett.129.270501}.

In this article, we present different low-dimensional tensor network structures and characterize them in terms of contraction costs and memory usage. 
We train them to copy random quantum states, which do not have any implicit correlation structure, 
using a set-up that relies on tensor contraction with a surrogate instead of training with samples (which allows us to control the complexity of the target system), as well as gradient-based tools with automatic differentiation.
In addition, we introduce the \textit{compact} version of any tree-TN structure and discuss when it can present an advantage. 

The results show that simpler structures achieve slightly better precisions only for the smallest bond dimensions, but those that are more dense have the advantage as the bond dimension grows. Specifically, we see that MPS and some large-diameter tree TNs perform worse than their more dense counterparts, even though the bond dimension is high enough to store the same amount of information. We also include some trainings with target systems of reduced complexity that are numerical evidence for the advantage of the introduced \textit{compact} tensor networks, and further support the findings on the impact of density.

Improving the efficiency in time and energy of HPC computations is a principal aspect of the field, driving the race for larger simulations. In recent years, GPU acceleration has become crucial to achieve these performance improvements \cite{menczer2024parallelimplementationdensitymatrix}, and has in fact been integrated in the latest generation Marenostrum 5 supercomputer. The training presented in this work has been built on libraries that allow us to compare the performance with and without acceleration. For these reasons, our results include a comparison between CPU-only nodes and GPU-accelerated nodes that shows an advantage in performance for large enough systems.

Finally, we discuss the relevance of these findings in the TN training field, and we conjecture that this behaviour is strongly related to the interplay between data distribution in the structure and training methods, supporting that DMRG and other TN native methods would indeed benefit from a TN structure imitating that of the target state. We recommend to view all plots in this article in colour to distinguish properly the data for each TN geometry.

\section{Tensor Network structure}\label{sec:structure}

\subsection{Bond dimension and correlations}

Entanglement is a crucial property of quantum systems that distinguishes them from systems with only classical properties \cite{entanglement_correlations}. It is related to the correlations between different parts of the system, and while it has been extensively studied in the case of bipartitions, characterizing and quantifying multipartite entanglement in general is still a profoundly complex topic \cite{horodecki, horodecki2024multipartiteentanglement}. 
For a quantum state $\ket{\psi}$ with density matrix $\rho = \ket{\psi}\bra{\psi}$, the subsystems A and B of a physical state $\rho$ are not entangled (and thus completely uncorrelated) when they can be written down as a tensor product $\rho = \rho_A \otimes \rho_B$ \cite{Nielsen_Chuang_2010}. Otherwise, this decomposition requires a sum of different product states, and $A$ exhibits entanglement with $B$.
This can be related to the Schmidt decomposition of a quantum state \cite{Nielsen_Chuang_2010}, which also for a bipartition $A/B$ is
\begin{equation}
\label{eq:schmidt}
    \ket{\psi} = \sum_i^{\chi_S} \lambda_i \ket{\phi_i}_A\otimes\ket{\phi_i}_B.
\end{equation}
When looking for the tensor product decomposition on $A/B$ of the density matrix of such a state, $\rho=\ket{\psi}\bra{\psi}$, it is clear that the amount of terms in the Schmidt decomposition dictates how many tensor product terms are needed.
We call the amount of states needed in this sum the Schmidt number $\chi_S$. Sometimes we don't need many elements in the decomposition, so that $\chi_S$ is small, which can happen for example when the entanglement between $A$ and $B$ is small.

Tensor networks \cite{orusPracticalIntroductionTensor2014} can encode high-dimensional tensors into the product of smaller, low-dimensional ones. This can be generally useful for data encoding \cite{9607509,su2024languagemodelingusingtensor} in the classical setting, where they are sometimes known as tensor trains \cite{tensortrain}, although they are more commonly used in simulation of quantum systems \cite{Orus_2019}. In the quantum setting, the high-dimensional tensor is commonly a quantum state of $n$ sites, which before decomposing needs to store $d^n$ coefficients, with $d$ the dimensionality of each site. For example, quantum digital computing uses qubits with $d=2$. Such a decomposition is advantageous whenever correlations between tensors that are far apart in the network are not too high, as it means that those sites present a small amount of entanglement. Then, a bipartition separating these tensors, following Eq. \ref{eq:schmidt}, does not need a large dimension $\chi_S$ to encode the full state, and the sum of memory required by all the small tensors can be smaller than that of the large, contracted tensor. The low-dimensional tensors can be connected through bonds in various geometries, which can be useful depending on the entanglement structure of the system they encode, although networks with more complexity and expressive power will entail higher performance costs.


One of the advantages of tensor networks is the flexibility of their graphic notation. Each tensor is represented by a small node, usually round-shaped, and the bonds between tensors are lines that connect these nodes. These are called virtual bonds, or virtual indices, as they depend on the geometry of the tensor network and not necessarily the properties of the system they encode. If their dimension is as large as the Schmidt number $\chi_S$ for the bipartition of the two parts of the system they connect, the tensor network is a faithful encoding of the system. We denote the maximum of the dimensions among all virtual bonds as the bond dimension $\chi$. The dimension of the system, on the other hand, is encoded in the physical bonds (indices), which are connected to a node on one side and free on the other. A few examples can be seen in Figure~\ref{fig:structures}.
The original high-dimensional tensor can be recovered by contracting all virtual bonds. 


A tensor network diagram is directly equivalent to an equation between the original tensor and the small tensors, which makes the contractions explicit. We use an MPS \cite{orusPracticalIntroductionTensor2014} example to illustrate this, which is a 1-dimensional chain and thus the simplest example of a non-trivial tensor network. Here, the high-dimensional tensor $T$ with $n$ physical bonds is equivalent to the contraction of $n$ tensors as:
\begin{equation}\label{eq:basic_tn}
    \mathcal{T}^{i_1 \dots i_n} = \sum_{k_1\dots k_{n-1}} (T_1)^{i_1}_{k_1}(T_2)^{i_2 k_1}_{k_2} \dots (T_N)^{i_n k_{n-1}}
\end{equation}
It is illustrative to see in this example how the bond dimension $\chi$ is linked to the entanglement properties of physical systems. For a separable state, any bipartition will have $\chi_S=1$, so it can be encoded into a TN with $\chi=1$. Other states can have entanglement limited to a 1D nearest-neighbours, such as the AKLT state \cite{affleckRigorousResultsValencebond1987}, which needs $\chi=2$. A maximally entangled state requires up to $\chi=2^{n/2}$, which is in fact the upper bound of $\chi_{S}$ for any system of $n$ qubits, so an MPS with this $\chi$ can represent any state with perfect precision. Beyond these exact cases, the bond dimension can also be reduced at the cost of precision. For example, even if a bipartition on a state has a large $\chi_S$, it can still be that most of the coefficients $\lambda_i$ are really small, which is equivalent to a low entropy $S$ across that bipartition. Then, a tensor network with $\chi<\chi_S$ can encode an approximate state $\ket{\psi_{TN}}$ that is almost identical to the original in terms of the fidelity $\braket{\psi|\psi_{TN}} \sim 1$.

Efficient simulations with tensor networks rely on limited entanglement, mostly between close neighbors \cite{ciracMatrixProductStates2021}, or on a hierarchical structure of entanglement \cite{vidalEntanglementRenormalization2007,evenblyTensorNetworkStates2011}. In the extreme case of no entanglement, the simulation cost of a system grows linearly with its size; in more complex cases, systems can adhere to an area law \cite{eisertAreaLawsEntanglement2010} that allows TN methods like DMRG \cite{mccullochInfiniteSizeDensity2008d} to perform efficient simulations with great success. 
Precisely because bond dimension makes entanglement and correlations in general both visible and accessible, tensor networks are very useful as a tool to understand quantum states, phases, and algorithms \cite{Orus_2019}. When we find that a quantum family of states can be simulated with a TN ansatz, we are obtaining a great deal of information about its correlation structure, and we can deduce that a system presents linear entanglement, area law entanglement, etc. Conversely, if we find a system with these properties, we know that TN ansätze are a good fit for simulation.

\subsection{Choice of structures}\label{subsec:structures}

\begin{figure}[t]
     \centering
     \includegraphics[width=1.02\linewidth]{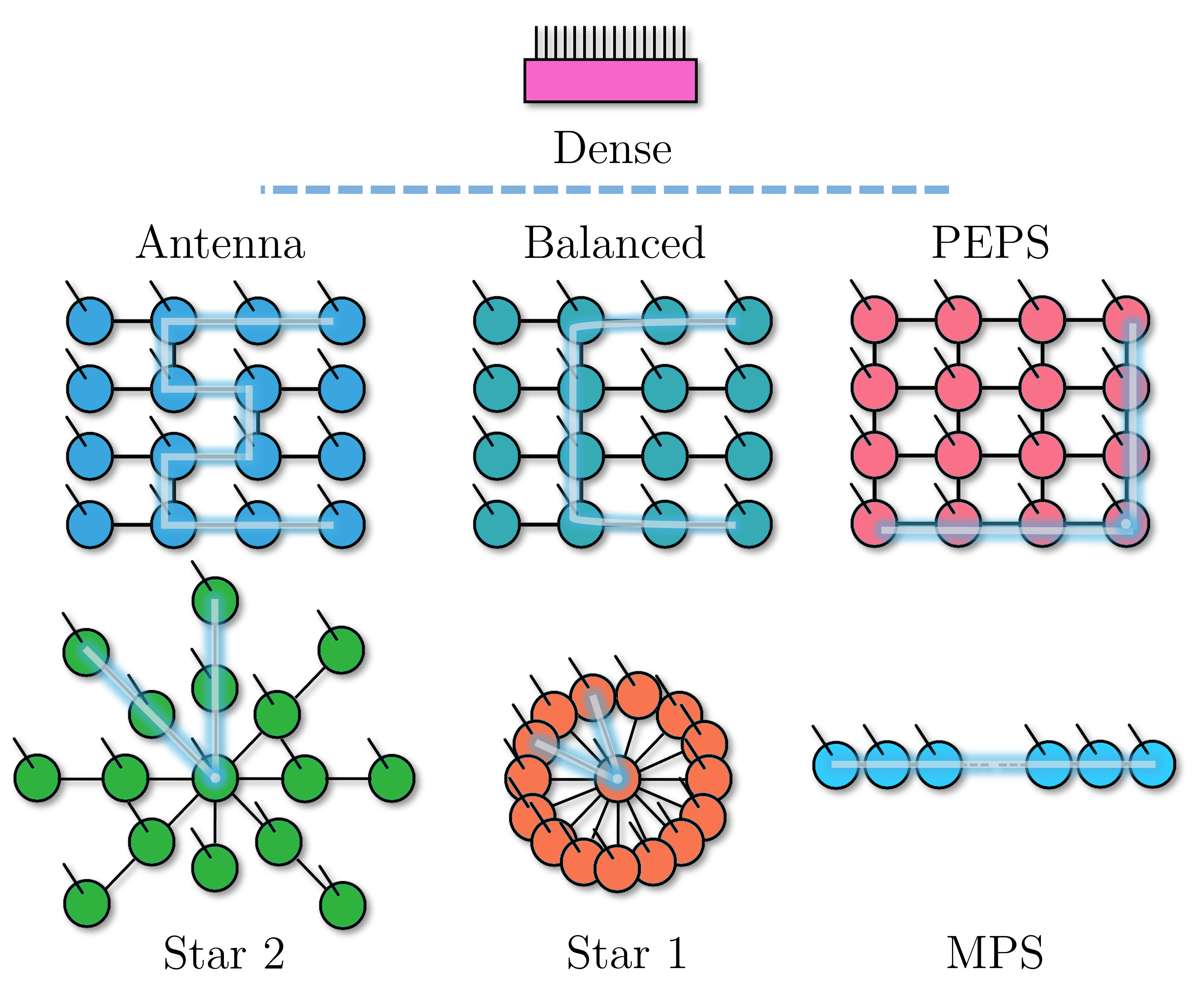}
     \caption{TN geometries used in this work, with a few tree-like TN (including MPS) and the PEPS ansatz. The highlight indicates the longest path between tensors, which we use as a measure of density. For PEPS, instead it shows the longest \textit{minimal} path between nodes when considering all node combinations. At the top, a single tensor equivalent to the contraction of any of the other geometries.
     } 
     \label{fig:structures}     
\end{figure}

We focus our study of gradient based tensor network training mostly on different tree tensor networks, including also the 2D projected entangled pair states (PEPS) \cite{verstraete2004renormalizationalgorithmsquantummanybody} geometry comparison when the size of the system allows it. In Fig. \ref{fig:structures} we show each different geometry used in the training. Beyond the known ansätze of MPS and PEPS, we have an ``Antenna" structure that reduces the distance between sites with respect to an MPS, without allowing tensors with more than 3 virtual indices, whereas``Balanced" achieves an even smaller distance at the cost of having bigger tensors with 4 virtual indices. We also propose a \textit{highly connected} tree with ``Star$_n$", where a single site is connected to many sites. For $n=1$, the site distance is minimal without introducing loops, and the central tensor is as complex as the \textit{dense} tensor, but the contractions of tensors on a single site is minimally expensive (comparable to the edge of an MPS). As $n$ grows, the beams of the star have more nodes and the distance between nodes increases. In the trainings we use $n=1,2,3$ (despite $n=3$ not being depicted in Fig. \ref{fig:structures}), and $n=1,2$ specially require much bigger tensors than the other tree structures.

Tree tensor networks have traditionally been used with the binary tree structure \cite{PhysRevA.74.022320}, which was also a precursor to the success of the multi-scale entanglement renormalization ansatz (MERA) \cite{PhysRevLett.99.220405}. This structure can be made more flexible by dropping the binary requirement \cite{10.1093/ptep/ptad018}. The structures that we study here are equivalent to this generalized tree TN but with some of the branches contracted. One can also allow for loops in the connectivity, which increases the representation power of the structure, but has to forgo most of the performance in the algorithms that make MPS very efficient (for example, using canonical tensors becomes much more complicated \cite{haghshenas2019conversion,hyatt2020dmrgapproachoptimizingtwodimensional}). Connectivity of the structure, once loops are allowed, can be taken beyond 1D MPS and 2D PEPS into the extreme all-to-all case, in which an $n$-site system would be encoded into an $n$-dimensional tensor network, reaching maximal contraction complexity as well as reprentation power. On the other hand, restricting the ansatz to be loop-free makes it possible to establish an ordering inside the structure, which allows MPS tools to be used while still making use of the increased flexibility beyond a chain structure \cite{PhysRevA.74.022320}.

Irregular tree structures such as the ones studied here are not very commonly used. The reason is that when optimizing systems that are not well characterized, a structure that works well in general and has highly developed tools becomes easy to use, and can also perform well despite not being optimal. However, an efficient heuristic that tailors a structure to a given problem could offer an additional advantage. Efforts in this direction have been made recently \cite{PhysRevResearch.5.013031}, with good results in terms of identifying entanglement structures in a specific system \cite{hikihara2024visualizationentanglementgeometrystructural}. These structures are thus a good fit for systems whose known properties picture a certain entanglement structure, from which one can propose a physically motivated ansatz. We contribute to this application by studying the benefits and drawbacks of the resulting structures in regards to training methods.

\subsection{Training with a surrogate system}\label{subsec:surrogate}

\begin{figure}[t]
     \centering
     \includegraphics[width=1.02\linewidth]{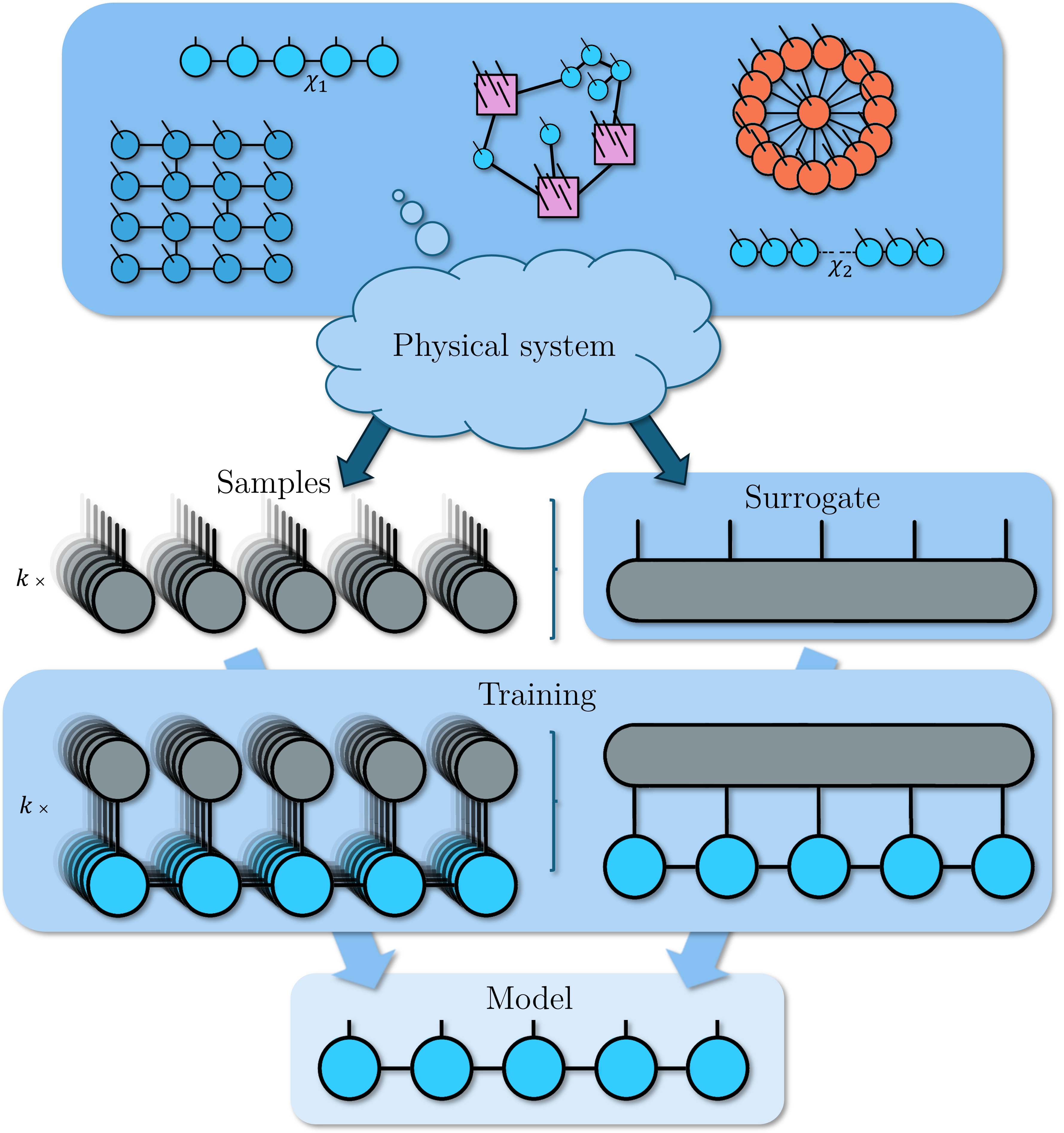}
     \caption{Usual training process of a TN model using samples of a target unknown physical system, to the left, compared to our method with a surrogate TN, to the right, where we control the amount of correlations in a TN with its bond dimension, and then we contract it into a single dense tensor. Both approaches output a model of the physical system in the form of another TN (in the example, an MPS).} 
     \label{fig:concept}     
\end{figure}

Training a model in machine learning usually requires a large amount of samples \cite{lu2023surveymachinelearningsamples} which contain the information that we want to learn. Thus, one can expect a training to consist of a set of $k$ samples $\phi_i$, which are extracted from a target physical system, and a model $\psi$ \cite{Fuksa_2024}, which is expected to generalize well following some known properties or theorems. In the setting of using tensor networks as the model $\psi$, this translates to a contraction between samples and the TN ansatz, which is commonly a simple MPS \cite{stoudenmire2017} or a similar structure \cite{TNAD_paper}. These contractions are accounted for as an ensemble in some appropriate cost function $\mathcal{L}(\phi_i,\psi)$. Such an approach must deal with issues like generalization of the model, sampling complexity, and so on, which are quite complex \cite{Cerezo_2022} and depend not only on the structure of the TN ansatz but also the training set and the properties of the model itself we are trying to learn. We can avoid some of these effects by using a surrogate TN that represents the physical system directly.

With the single tensor surrogate method, depicted in Fig.~\ref{fig:concept}, we can prepare physical systems that have specific entanglement structures or limited correlations by initializing a random tensor network with a given structure ansatz, fixing $\chi$, and then contracting it into a single tensor. In this final statevector-like form, the surrogate \textit{hides} its origin, as it would happen for a target physical system with unknown properties. Then, instead of many contractions with each sample $\phi_i$, we perform a single contraction between the surrogate and the model which avoids the statistical problems of sampling but, on the other hand, is more costly computationally. The final trained model, in the ideal case of perfect training, should be equivalent to the physical system itself. Since the model has a tensor network form, it is then equivalent to the surrogate itself. In this work we focus on controlling the complexity of the target system (which would extend to that of the samples) via the bond dimension $\chi$ of the surrogate for two different scenarios, as we will explain in Section~\ref{sec:simulation}, and leave for future work the study of encoding different structures into the surrogate to see if we can recover them.

The structure of the tensor network model plays a crucial role in several ways. It directly constrains how hard it is, computationally, to contract the network. Consequently, it also characterizes how hard it is to find the optimal path for such contraction, with an MPS on one end (the optimal path is known beforehand) and the contraction of non-structured tensor networks, like in arbitrary quantum circuits, \cite{Gray_2024} on the other (known heuristics can find good paths, but not optimal). It also relates to how well they can represent the system that is being encoded in them. This is an obstacle that can be overcome with enough resources, as we have seen that without binding $\chi$ even an MPS could represent the most complex quantum state. However, we will see that even with large $\chi$, wide geometries that have enough correlations between sites can still fail to train properly. More generally, finding the appropriate geometry that reflects the correlations between different parts of the target quantum system can keep bond dimension costs lower, and thus the computational cost, but such benefits must compete with how easy it is to find and train them.

\subsection{Tensor budget}

\begin{figure*}[ht]
     \centering
     \includegraphics[width=0.9\linewidth]{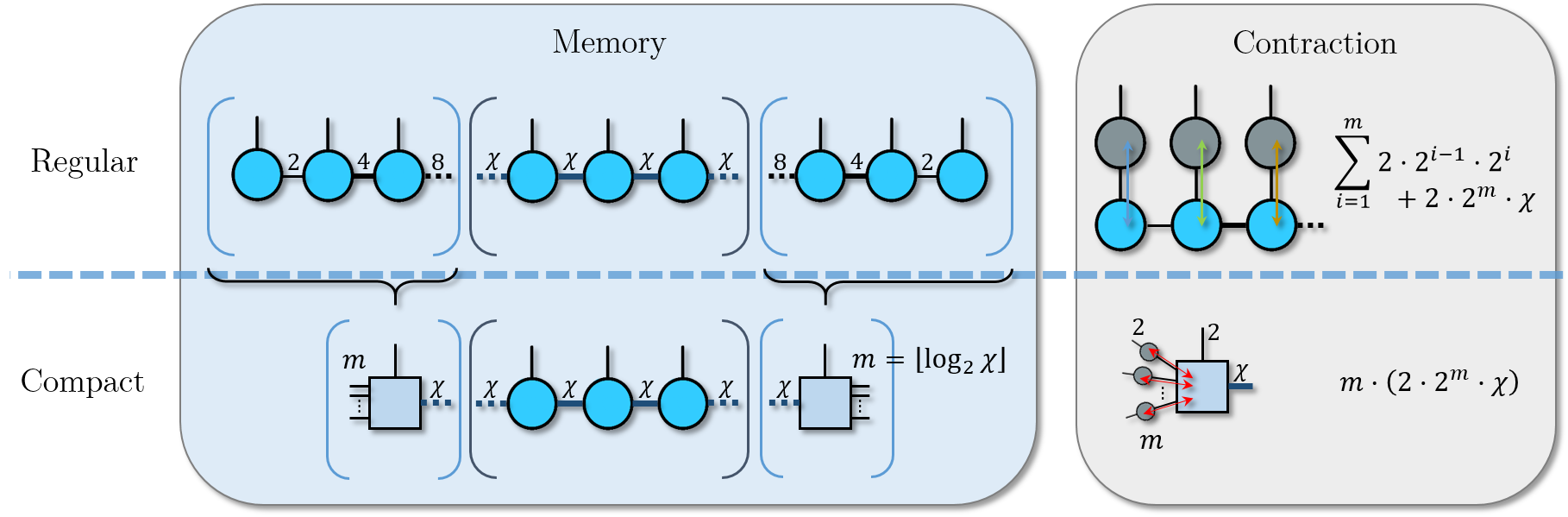}
     \caption{Example using an MPS of the ``compact" TN approach, where bonds smaller than $\chi$ are contracted. This reduces the memory needed to store the tensor network, but increases the cost of the contraction for the compacted sites.
     } 
     \label{fig:compact}     
\end{figure*}

In the design of our tensor network, it is important to budget the resources properly. Memory limitations are very relevant in the size of simulations that can be achieved \cite{wu2018memoryefficientquantumcircuitsimulation,pan2024efficientquantumcircuitsimulation}, and common strategies to improve performance trade between memory and computational cost, especially in the HPC setting \cite{Sanchez_Ramirez_2021}. This tradeoff has a direct equivalent in tensor networks. In an MPS with a fixed bond dimension $\chi$, for example, we can contract all sites with a virtual index smaller than $\chi$ without losing information or correlations. This is evident when using the maximum $\chi=\chi_S$: for an odd $n$-qubit system, the central tensor with one physical bond and two virtual bonds has the same dimension ($2^{(n-1)/2)}\times2\times2^{(n-1)/2)}$) as a dense representation of the same quantum system ($2^{n}$). On the other hand, each contraction on these \textit{redundant} sites becomes more expensive, which is relevant for TEBD \cite{PhysRevLett.93.040502,PhysRevB.102.035115} or circuit simulation \cite{pan2024efficientquantumcircuitsimulation}, as depicted in Fig. \ref{fig:compact}. This method can be trivially generalized to any TN structure without loop. Such structures employ more tensors than strictly necessary for the amount of information that they store; from another optic, the information is distributed over a larger amount of numbers that need to be trained. 

We can show explicitly when this contraction effect is relevant in MPS. For a given $\chi$, and assuming that the physical indices have fixed dimension $p$, let us consider two tensors that share a bond with dimension $d<\chi$. The tensor that is further from the center of the MPS has a left virtual index of dimension at most $d/p$, so in total it needs a memory at most $d^2$, whereas the other tensor needs $d\cdot p \cdot d_r$, where $d_r$ is its other virtual bond. The contracted tensor then occupies a memory of $d/p \cdot p \cdot p \cdot d_r$ which is the same as the second tensor. Thus, the sum of both uncontracted tensors will always be strictly larger. Starting from one end of the MPS, if we do this $m$ times before reaching the tensor with virtual bond dimension $\chi$, we will have a tensor with dimensions $p^m \cdot \chi$. Therefore, $m$ is the largest integer smaller than $\log_p \chi$. In particular, for the usual case where $p=2$ we have $m=\left \lfloor{\log_2\chi}\right \rfloor$. We can repeat this process from each end of the MPS.

The closed formula for the MPS case cannot be generalized to arbitrary tree TNs, as it depends on the branching structure, but we can apply the same concept to decide which tensors to contract. Considering the leaves of the tree as the tensors with only one physical bond and one virtual bond, we will start contracting the branches until we reach a tensor with a bond of size $\chi$. For tree tensor networks with few branches (or where most branches connect to few tensors) the effect will be similar to that of an MPS, whereas for many branches the memory reduction will be much more substantial.

This effect of the compactification is rarely the leading performance or memory factor in large MPS simulations, since that depends on the large tensors that are far from the endpoints (leaves) of the network. Specifically, it is negligible when $n>>2\left \lfloor{\log_2\chi}\right \rfloor$. Regardless, it can be used to increase the quality of the results that are achievable with a fixed memory budget, especially with low bond dimensions or sites. On the other hand, tree structures can benefit much more, namely those with small average distance between nodes as they have a larger amount of leaves. If the distance is too short or the bond dimension too large, this scheme naturally outputs a single dense tensor (i.e. the statevector, for quantum systems). This only happens when the resources are high enough to compute the whole system, in which case training the dense tensor is indeed the optimal approach when the target has close to maximal complexity, as we show in this article. To quantify the effect of this modification at a practical level, we use a target system with reduced complexity and train a \textit{compact} version of each tree structure in Fig. \ref{fig:structures}. This way, we can achieve good simulations with a smaller bond dimension and see the effects of this technique without having all compact TN be equivalent to the dense tensor. As in the rest of simulations, we study the training in terms of time, contraction cost, and memory usage.

\subsection{Barren Plateaus}\label{sec:plateau}

Barren plateaus are an obstacle in the training of current quantum variational algorithms \cite{McClean_2018}, and their existence has recently been linked to simulability \cite{cerezo2024doesprovableabsencebarren} for all known trainable instances, supporting previous suggestions of a structural rethinking in the current search for quantum advantage in quantum machine learning \cite{Schuld_2022}. In the study of TN for computation, however, this outlook plays an opposite role. It motivates the exploitation and improvement of TN with quantum algorithms to define the limit of quantum advantage by pushing the boundary of simulation techniques. 

The influence of this problem extends to the training of tensor networks, as has been shown in the last years \cite{PhysRevLett.129.270501} -- perhaps unsurprisingly considering the connection of the structures to quantum computing. The structure of the TN in use affects greatly the appearance of plateaus  \cite{basheer2024trainabilityclassicalsimulabilitylearning, Cervero_Mart_n_2023}, with MPS being the most susceptible. This effect should be considered together with our findings on memory, efficiency and precision when judging our conclusions on the suitability between structures and training applications. In fact, for certain sizes of the system we encounter them in our training of MPS despite being absent for the denser geometries, as we will show in the results section. While this is consistent with previous findings, the progressive improvement in success rate as the density increases shows that there is not a categorical distinction between MPS and the rest of structures; instead, there is a somewhat continuous correlation between the complexity of the TN and the appearance of barren plateaus.

\section{Simulation}\label{sec:simulation}

The training algorithm we designed encodes random states with controlled entanglement on a single tensor (see Sec. \ref{subsec:surrogate} and Fig.~\ref{fig:concept}), which we use to train each of the proposed geometries (see Sec.~\ref{subsec:structures}) to match the random state. We employ a cost function $\mathcal{L}$ based on the infidelity of a given state with the target. This is defined as
\begin{equation}\label{eq:infidelity}
\mathcal{I} = 1- \mathcal{F} = 1- \braket{\Psi_{surrogate}|\Psi},
\end{equation}
whereas the loss function that we chose is
\begin{equation}\label{eq:loss}
\mathcal{L}(\Psi) = (\log(\mathcal{F})-1)^2.
\end{equation}
Simpler functions on infidelity such as $\mathcal{L} = \mathcal{I}^2 = (1-\mathcal{F})^2$ were tested with worse results. For the optimization, we used L-BFGS-B, a pseudo-second order gradient descent approach that only calculates the function and its gradient, but uses them to approximate the Hessian matrix to guide the training with a higher derivative while avoiding its expensive calculation. The quality of results was compared to other similarly powerful optimization methods with little to no difference, while L-BFGS-B performed slightly faster for the small sizes of the problems on which we tested the optimizer, hence our choice.

Gradients were calculated using automatic differentiation (AD), a generalization to backpropagation, which as a tool has had a great impact in the field of machine learning and has since found applications in other areas~\cite{baydin2018automaticdifferentiationmachinelearning}. It can avoid some pitfalls of symbolic differentiation algorithms without resorting to numerical methods. For any given function, AD maps its dependency on other functions all the way down to elementary operations (whose derivatives are well known) by going through the code, and can evaluate it at any given point. Our choice of the library to manage tensor networks, QUIMB \cite{Gray2018}, is compatible with different AD implementations. We used JAX \cite{jax2018github}, which also offers integration with GPUs and certain parallelization tools. After the training, we focus on different traits about the trained geometry, namely the largest tensor, the size of the full TN (sum of all its tensor sizes), its bond dimension $\chi$ or the optimization time, to compare to the quality of the learning, measured with infidelity.

The computation was performed with \textit{Marenostrum 5}~\cite{mn5} which has "non-accelerated" nodes of up to 112 cores (powered with two Intel Sapphire Rapids 8480+ with 56 cores at 2Ghz), 256 GB of DDR5 memory, as well as accelerated nodes of up to 80 cores per node (powered by two Intel Sapphire Rapids 8460Y+ with 40 cores at), 512 GB of DDR5 memory and 4x Nvidia Hopper GPU with 64 HBM2 memory. Due to the scale of our simulations, we compare the optimization time using a limited amount of cpu cores against a node using a single GPU, leaving a full scale implementation on multiple nodes for future work, with bigger sizes and a selection of tensor network structures based on the outlook of this work. We show in Sec. \ref{sec:results} that the accelerated nodes improved the performance of the training for some sizes. It is also relevant that GPUs are often used in settings where there is no need for a large numerical precision, which contributes to their performance. However, we found that the precision of our calculations was affected by this, and only with higher precisions were the infidelity results at the level of CPU only runs. Thus, performance gains of using GPU for tensor network contraction appear smaller than expected due to the specific nature of the simulations.

\begin{figure}[b]
     \centering
     \includegraphics[width=\linewidth]{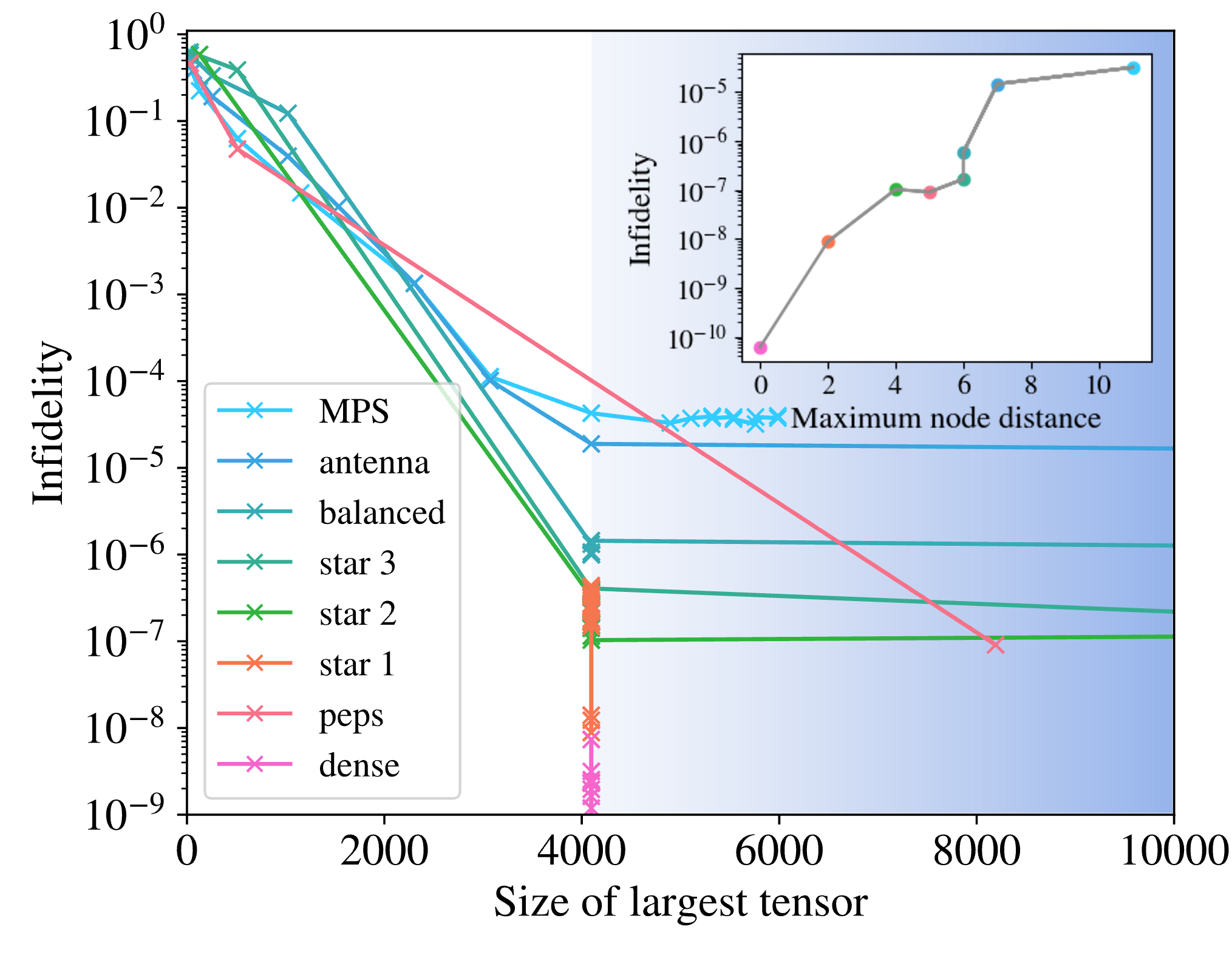}
     \caption{Training performance measured with infidelity as a function of largest tensor in the TN, for $n=12$ sites. The size of the tensors as controlled with bond dimension for each of the geometries in \ref{fig:structures}. Trainings in the blue shaded area are using a bond dimension $\chi>\chi_S$. In the inset, the best infidelity reached against the maximum node distance of the TN structure. The ordering follows that of the legend.
     } 
     \label{fig:main_plot}     
\end{figure}

\section{Results}\label{sec:results}

The training described in the previous sections reveals that the connectivity of the tensor network structure directly affects the maximum precision that they can reach. In Figure~\ref{fig:main_plot} we show how the infidelities for simulations of a system of $n=12$ sites approach $0$, and thus improve, as we increase the bond dimension of the structure. They plateau when the maximal bond dimension $\chi_{S}$ is reached, which is to be expected from the point of view of complexity of the quantum state, but it proves that the redundant information does not help the training procedure. In the inset, we can see how this effect correlates with the maximum distance between nodes in the structure, as highlighted in Fig.~\ref{fig:structures}. This means that structures that are more dense, in the sense of proximity between nodes, perform better in the training. This is the case even if the bond dimension is high enough to represent the target state perfectly ($\chi=\chi_S$), for which ideally the infidelity should decrease to machine precision for any structure. We also see that increasing the amount of information that the TN can store, which we do by using bond dimensions $\chi$ beyond $\chi_S$, does not help the training and, in fact, hurts the success rate as we see below. For PEPS, we observe that the first data point with good convergence happens for a large ``\textit{size of largest tensor}". This is due to the structure of PEPS, as the largest tensor grows as $\chi^4$. Since bond dimension must be an integer, this number jumps from being small to $\chi^4>2^n$. For a tensor in a tree structure, each bond dimensions $\chi$ is bounded by the amount of nodes they connect to, meaning that the largest tensor \textit{is} bounded by $2^n$, explaining the vertical alignment of the points.

\begin{figure}[t]
     \centering
     \includegraphics[width=\linewidth]{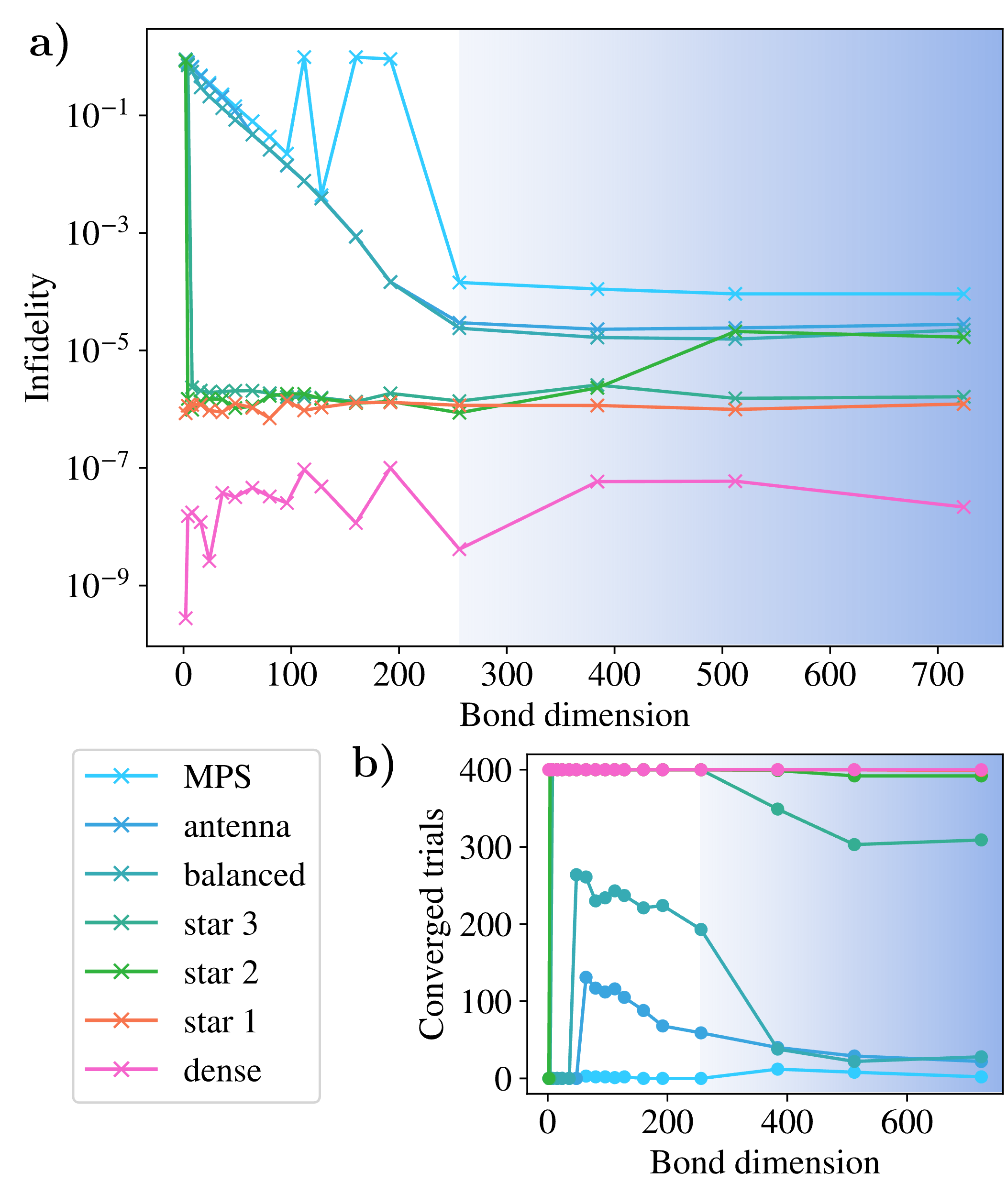}
     \caption{Training performance measured with infidelity as a function of bond dimension, for $n=16$ sites, in a).  For some bond dimensions in MPS, none of the trainings reach a good infidelity. The amount of trainings that cross a $10^{-3}$ threshold for infidelity is plotted in b), showing a decrease with the amount of information in the structure, as measure by bond dimension $\chi$. The blue shaded area indicates $\chi>\chi_S$.
     } 
     \label{fig:plateau_plot}     
\end{figure}

In Figure~\ref{fig:plateau_plot} we show a training with larger geometries ($n=16$) and plot the infidelities against bond dimension. The denser structures achieve high fidelity for low bond dimensions, whereas those that are less dense need to reach $\chi=\chi_S$. Beyond underlining the previous conclusions on infidelity vs structure density, we can see that even with $400$ trials the MPS training fails to converge, showing its shortcomings in this kind of training and agreeing with current research around barren plateaus (discussed in section~\ref{sec:plateau}). The convergence of the training with $\chi=128$ and $\chi\geq 256$ shows that this is not due to the structure not being able to store enough information, but to the training process. In Fig.~\ref{fig:plateau_plot}b we plot the amount of trainings that cross the threshold of $\mathcal{I}=10^{-3}$, seeing that they decrease with bond dimension. For the most powerful structures (dense, star1 and star2), the decrease is not meaningful. For the rest, the decrease is more pronounced in the regime where $\chi>\chi_{S}=256$, showing that the additional information in the structure makes the training harder even though, as seen above, this does not translate into an improved infidelity.

\begin{figure}[b]
     \centering
     \includegraphics[width=1.04\linewidth]{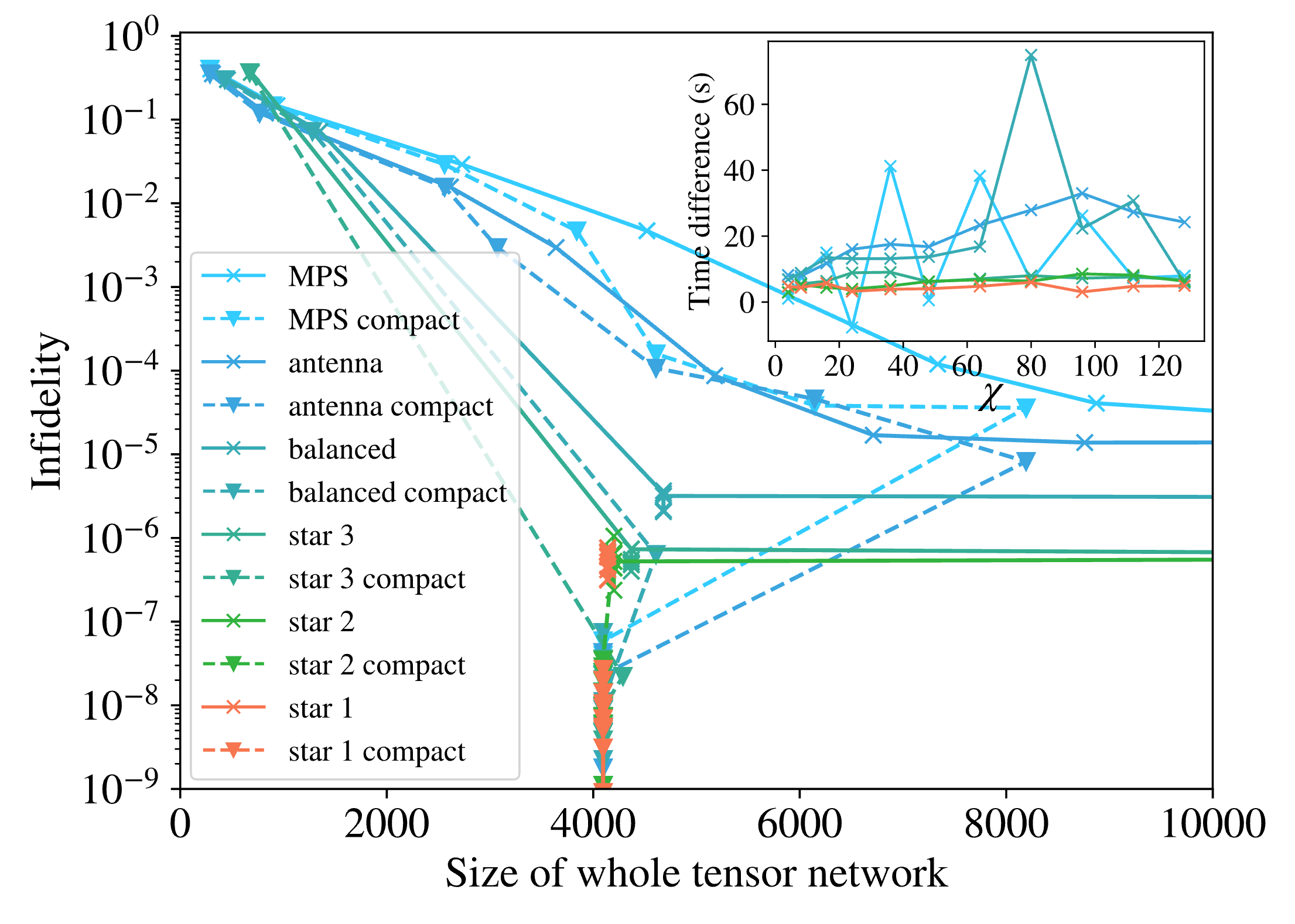}
     \caption{Training performance measured with infidelity as a function of the full size of the tensor network for the examples in Fig.~\ref{fig:structures}, and the compact version of each geometry that follows from Fig.~\ref{fig:compact}. The last data points for the compact TNs decrease in size because the compact threshold is high enough to contract them into a single tensor.
     } 
     \label{fig:compact_plot}     
\end{figure}

The compactification approach shown in Fig.~$\ref{fig:compact}$ is not shown in the previous plots, as doing so with $\chi=\chi_S$ leads to a single large tensor and there would be no difference between structures. To visualize the effect we need to test it with reduced bond dimensions, but for random quantum states this would lead to very high infidelities. We can, however, use the hidden data structure of Fig.~\ref{fig:concept} to prepare targets with reduced complexity equivalent to a limited $\chi$, and perform the training using that same $\chi<\chi_S$. In Fig.~\ref{fig:compact_plot} we show the results of such training and see that for each geometry, the corresponding compact version trains better by either achieving the same precision at smaller bond dimension or reaching lower infidelities at the same bond dimension. This is visible as the dashed lines (for compact TN) always run closer to the origin, below the solid lines (for regular TN) in the y axis and to the left in the x axis. We see that the last points for each line, corresponding to the biggest $\chi$, collapse to the area where the dense trainings lie, because the threshold $\chi$ for the compactification is high enough to reduce them to a single tensor.



\begin{figure}[t]
     \centering
     \includegraphics[width=\linewidth]{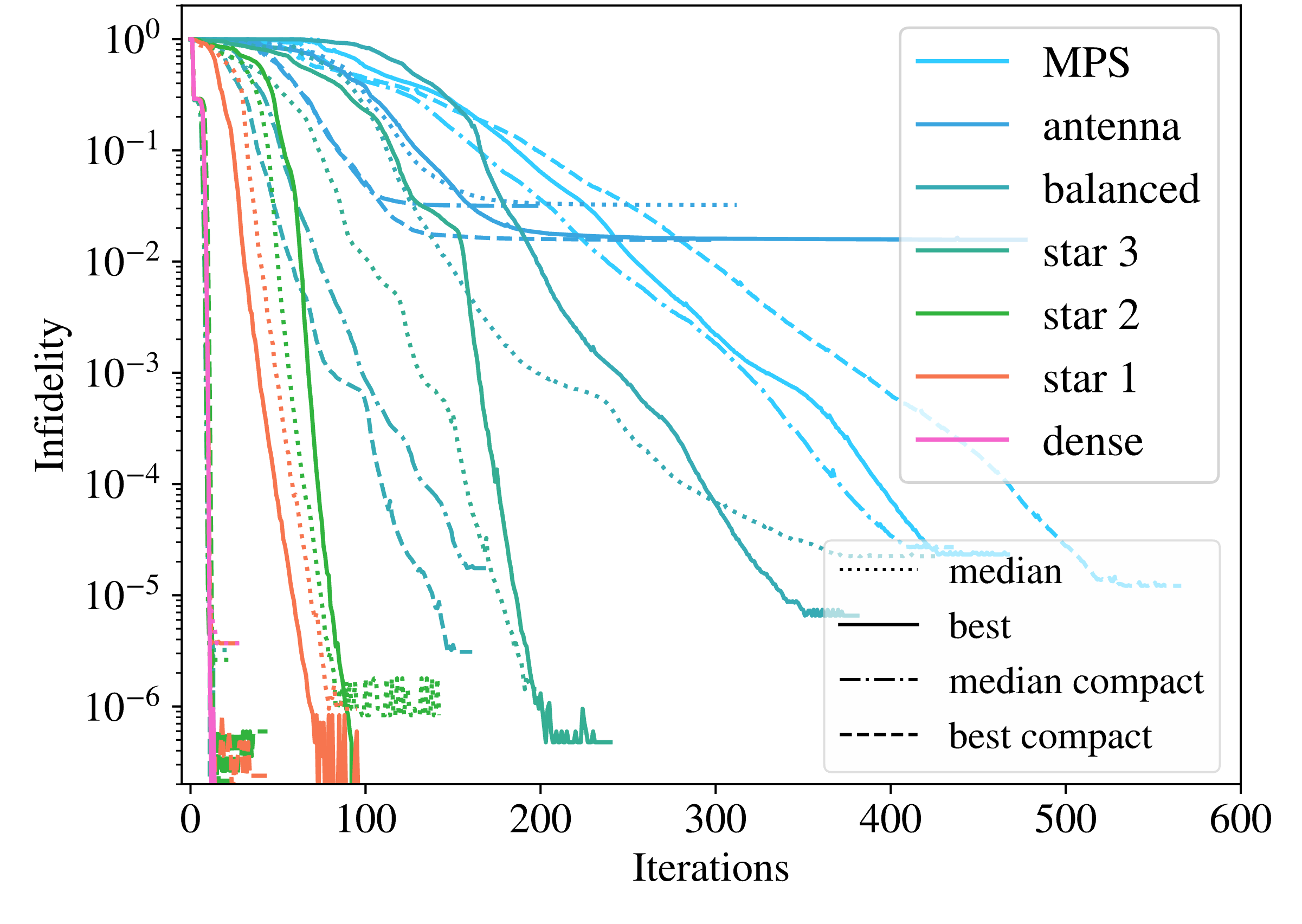}
     \caption{Example of training with $n=16$ sites, for each geometry and using low bond dimension to compare with the compact tensor network structures. We distinguish between solid (dashed) lines for the best regular (compact) training and dotted (dot-dashed) for the regular (compact) median training, ordered by final infidelity. 
     } 
     \label{fig:training_compact}     
\end{figure}

An example of the training evolution across iterations is shown in Fig.~\ref{fig:training_compact}, for $n=16$, highlighting differences in the behaviour of all studied geometries. Across a run of $100$ trainings, we include both the best training and the median training, ordered by the final infidelity reached, as the difference in orders of magnitude skews the arithmetic mean. Results indicate that the compact version of geometries train faster and better than their regular counterparts, not only in the best case scenario but also on average. The improvement is consistent with our findings of smaller infidelity with growing density, as compactification reduces \textit{maximal node distance} and other similarly related measures of graph connectivity that correlate with it. We also see that below infidelities of $10^{-6}$ there are some numerical stabilities that impact the denser structures, and were the cause for the high variance in the dense training of Fig.~\ref{fig:plateau_plot}.

\begin{figure}[t]
     \centering
     \includegraphics[width=\linewidth]{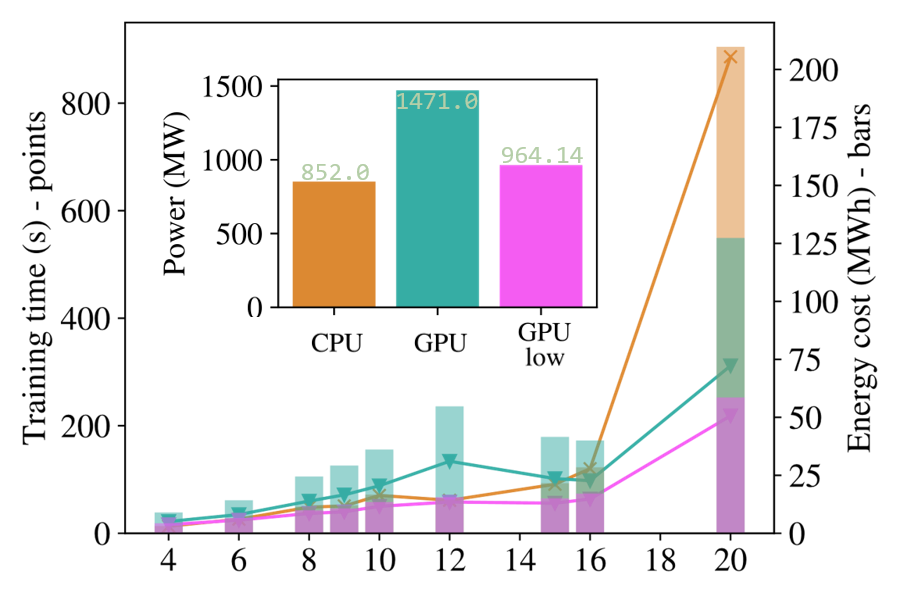}
     \caption{Time cost of the training in this work for different sizes of the system (left axis, points) and energy cost (right axis, bars), adding together all reviewed geometries. We use 20 CPU cores, and 20 CPU cores + GPU both with high and reduced precision, labeled as ``CPU", ``GPU" and ``GPU low" respectively. In the inset, the average power needed for each training.
     } 
     \label{fig:performance}     
\end{figure}

Lastly, we compare the impact of accelerated simulation on the time and energy efficiency of the training. The results in Fig.~\ref{fig:performance} compare a training using $20$ CPU cores against one using the same amount of cores in addition to $1$ GPU, with single and double precision in the GPU case. We can see an advantage in the GPU acceleration for the largest size $n=20$ when comparing trainings with the same precision (double). The default precision (single) when using a GPU with \textit{Jax} introduced some error in the results, leading to infidelities between 1 and 2 orders of magnitude larger than the main findings above. This could be good enough in some settings, and can be leveraged to achieve performance enhancements at smaller size, starting at $n=12$. In terms of energy, we see that GPU accelerated trainings draw around $50\%$ more power, which means that the total energy cost (power $\cdot$ time) is smaller when the training time is smaller by $1.5$, as is the case for $n=20$. Additionally, we identified that only part of the optimization was able to effectively use the GPU, reducing the impact of the advantage and contributing to the time advantage becoming apparent only for the largest sizes.

\section{Conclusions and outlook}\label{sec:outlook}

We have shown that different structures can greatly affect the training procedure of TN learning schemes in the context of machine learning. Crucially, more dense distributions of information can reach smaller errors despite the increased contraction cost, and even do so faster both in time and iterations. This means the way the TN structure connects the information that it stores is much more important than the total amount of information it can store, as evidenced by the worse performance of MPS and some tree TN with maximal bond dimension and further supported by the lack of improvement when allowing larger bond dimensions than necessary.

The emergence of barren plateaus in this type of gradient training for TN is relevant as we use a global loss function based on the infidelity with the target system. It has previously been linked to MPS, with tree TN and other more complex structures being less prone to exhibiting them. We have expanded on these findings by showing that their behaviour is tied to the density of the structure, with barren plateaus being absent for very dense structures and getting slightly more prominent as sparsity increases. For MPS, specifically, the training even fails to converge for $n=16$ sites unless using hundreds of trials.

Memory limits in the training of tensor networks are dependent on the largest tensor contraction at any point by relying on the transfer of large tensors from fast to slow memory. Regardless, decreasing the total memory needed contributes to reducing the overall duration of the computation by minimizing the movements of large tensors between memories. We have introduced a compactification ot tree TN structures, using our training examples to prove that it not only reduces the total memory needed but also improves the quality and speed of the training. The latter is due to a reduced number of iterations, as the cost of the contractions themselves increase slightly with the compactification. Finally, we have showcased an application of existing tensor network libraries with HPC resources, and also characterised performance improvements with the use of accelerated hardware in this same setting.

This work contributes to better understanding the training of these structures, and we think it is of great interest to extend the study of these structures beyond gradient training towards powerful methods like DMRG or TEBD in the future. Similarly, with the proposed surrogate procedure, one can test if entanglement structures hidden in a single tensor can be learned with tensor networks, either using gradient methods or DMRG.

\acknowledgments

We want to thank Sergio Sánchez and Germán Navarro for fruitful discussions around tensor networks and HPC performance, Berta Casas for her eye for plots, and the rest of the BSC team for their support and comments. Authors acknowledge financial support from EU grant HORIZON-EIC-2022-PATHFINDEROPEN-01-101099697, QUADRATURE, from Generalitat de Catalunya grant 2021SGR00907, and funding from the Spanish Ministry for Digital Transformation and of Civil Service of the Spanish Government through the QUANTUM ENIA project call - Quantum Spain, EU through
the Recovery, Transformation and Resilience Plan – NextGenerationEU within the framework of the
Digital Spain 2026.

\bibliography{bib}

\begin{thebibliography}{52}%
\makeatletter
\providecommand \@ifxundefined [1]{%
 \@ifx{#1\undefined}
}%
\providecommand \@ifnum [1]{%
 \ifnum #1\expandafter \@firstoftwo
 \else \expandafter \@secondoftwo
 \fi
}%
\providecommand \@ifx [1]{%
 \ifx #1\expandafter \@firstoftwo
 \else \expandafter \@secondoftwo
 \fi
}%
\providecommand \natexlab [1]{#1}%
\providecommand \enquote  [1]{``#1''}%
\providecommand \bibnamefont  [1]{#1}%
\providecommand \bibfnamefont [1]{#1}%
\providecommand \citenamefont [1]{#1}%
\providecommand \href@noop [0]{\@secondoftwo}%
\providecommand \href [0]{\begingroup \@sanitize@url \@href}%
\providecommand \@href[1]{\@@startlink{#1}\@@href}%
\providecommand \@@href[1]{\endgroup#1\@@endlink}%
\providecommand \@sanitize@url [0]{\catcode `\\12\catcode `\$12\catcode
  `\&12\catcode `\#12\catcode `\^12\catcode `\_12\catcode `\%12\relax}%
\providecommand \@@startlink[1]{}%
\providecommand \@@endlink[0]{}%
\providecommand \url  [0]{\begingroup\@sanitize@url \@url }%
\providecommand \@url [1]{\endgroup\@href {#1}{\urlprefix }}%
\providecommand \urlprefix  [0]{URL }%
\providecommand \Eprint [0]{\href }%
\providecommand \doibase [0]{https://doi.org/}%
\providecommand \selectlanguage [0]{\@gobble}%
\providecommand \bibinfo  [0]{\@secondoftwo}%
\providecommand \bibfield  [0]{\@secondoftwo}%
\providecommand \translation [1]{[#1]}%
\providecommand \BibitemOpen [0]{}%
\providecommand \bibitemStop [0]{}%
\providecommand \bibitemNoStop [0]{.\EOS\space}%
\providecommand \EOS [0]{\spacefactor3000\relax}%
\providecommand \BibitemShut  [1]{\csname bibitem#1\endcsname}%
\let\auto@bib@innerbib\@empty
\bibitem [{\citenamefont {Cirac}\ \emph {et~al.}(2021)\citenamefont {Cirac},
  \citenamefont {Perez-Garcia}, \citenamefont {Schuch},\ and\ \citenamefont
  {Verstraete}}]{ciracMatrixProductStates2021}%
  \BibitemOpen
  \bibfield  {author} {\bibinfo {author} {\bibfnamefont {J.~I.}\ \bibnamefont
  {Cirac}}, \bibinfo {author} {\bibfnamefont {D.}~\bibnamefont {Perez-Garcia}},
  \bibinfo {author} {\bibfnamefont {N.}~\bibnamefont {Schuch}},\ and\ \bibinfo
  {author} {\bibfnamefont {F.}~\bibnamefont {Verstraete}},\ }\bibfield  {title}
  {\bibinfo {title} {Matrix {{Product States}} and {{Projected Entangled Pair
  States}}: {{Concepts}}, {{Symmetries}}, and {{Theorems}}},\ }\href
  {https://doi.org/10.1103/RevModPhys.93.045003} {\bibfield  {journal}
  {\bibinfo  {journal} {Reviews of Modern Physics}\ }\textbf {\bibinfo {volume}
  {93}},\ \bibinfo {pages} {045003} (\bibinfo {year} {2021})},\ \Eprint
  {https://arxiv.org/abs/2011.12127} {2011.12127} \BibitemShut {NoStop}%
\bibitem [{\citenamefont {Markov}\ and\ \citenamefont
  {Shi}(2008)}]{Markov_2008}%
  \BibitemOpen
  \bibfield  {author} {\bibinfo {author} {\bibfnamefont {I.~L.}\ \bibnamefont
  {Markov}}\ and\ \bibinfo {author} {\bibfnamefont {Y.}~\bibnamefont {Shi}},\
  }\bibfield  {title} {\bibinfo {title} {Simulating quantum computation by
  contracting tensor networks},\ }\href {https://doi.org/10.1137/050644756}
  {\bibfield  {journal} {\bibinfo  {journal} {SIAM Journal on Computing}\
  }\textbf {\bibinfo {volume} {38}},\ \bibinfo {pages} {963–981} (\bibinfo
  {year} {2008})}\BibitemShut {NoStop}%
\bibitem [{\citenamefont {Gray}\ and\ \citenamefont {Chan}(2024)}]{Gray_2024}%
  \BibitemOpen
  \bibfield  {author} {\bibinfo {author} {\bibfnamefont {J.}~\bibnamefont
  {Gray}}\ and\ \bibinfo {author} {\bibfnamefont {G.~K.-L.}\ \bibnamefont
  {Chan}},\ }\bibfield  {title} {\bibinfo {title} {Hyperoptimized approximate
  contraction of tensor networks with arbitrary geometry},\ }\bibfield
  {journal} {\bibinfo  {journal} {Physical Review X}\ }\textbf {\bibinfo
  {volume} {14}},\ \href {https://doi.org/10.1103/physrevx.14.011009}
  {10.1103/physrevx.14.011009} (\bibinfo {year} {2024})\BibitemShut {NoStop}%
\bibitem [{\citenamefont {Vidal}(2003)}]{Vidal_2003}%
  \BibitemOpen
  \bibfield  {author} {\bibinfo {author} {\bibfnamefont {G.}~\bibnamefont
  {Vidal}},\ }\bibfield  {title} {\bibinfo {title} {Efficient classical
  simulation of slightly entangled quantum computations},\ }\bibfield
  {journal} {\bibinfo  {journal} {Physical Review Letters}\ }\textbf {\bibinfo
  {volume} {91}},\ \href {https://doi.org/10.1103/physrevlett.91.147902}
  {10.1103/physrevlett.91.147902} (\bibinfo {year} {2003})\BibitemShut
  {NoStop}%
\bibitem [{\citenamefont
  {McCulloch}(2008)}]{mccullochInfiniteSizeDensity2008d}%
  \BibitemOpen
  \bibfield  {author} {\bibinfo {author} {\bibfnamefont {I.~P.}\ \bibnamefont
  {McCulloch}},\ }\href {http://arxiv.org/abs/0804.2509} {\bibinfo {title}
  {Infinite size density matrix renormalization group, revisited}} (\bibinfo
  {year} {2008}),\ \Eprint {https://arxiv.org/abs/0804.2509} {0804.2509}
  \BibitemShut {NoStop}%
\bibitem [{\citenamefont
  {Vidal}(2007{\natexlab{a}})}]{vidalEntanglementRenormalization2007}%
  \BibitemOpen
  \bibfield  {author} {\bibinfo {author} {\bibfnamefont {G.}~\bibnamefont
  {Vidal}},\ }\bibfield  {title} {\bibinfo {title} {Entanglement
  {{Renormalization}}},\ }\href {https://doi.org/10.1103/PhysRevLett.99.220405}
  {\bibfield  {journal} {\bibinfo  {journal} {Physical Review Letters}\
  }\textbf {\bibinfo {volume} {99}},\ \bibinfo {pages} {220405} (\bibinfo
  {year} {2007}{\natexlab{a}})}\BibitemShut {NoStop}%
\bibitem [{\citenamefont {Verstraete}\ and\ \citenamefont
  {Cirac}(2004)}]{verstraete2004renormalizationalgorithmsquantummanybody}%
  \BibitemOpen
  \bibfield  {author} {\bibinfo {author} {\bibfnamefont {F.}~\bibnamefont
  {Verstraete}}\ and\ \bibinfo {author} {\bibfnamefont {J.~I.}\ \bibnamefont
  {Cirac}},\ }\href {https://arxiv.org/abs/cond-mat/0407066} {\bibinfo {title}
  {Renormalization algorithms for quantum-many body systems in two and higher
  dimensions}} (\bibinfo {year} {2004}),\ \Eprint
  {https://arxiv.org/abs/cond-mat/0407066} {arXiv:cond-mat/0407066
  [cond-mat.str-el]} \BibitemShut {NoStop}%
\bibitem [{\citenamefont {Begušić}\ \emph {et~al.}(2024)\citenamefont
  {Begušić}, \citenamefont {Gray},\ and\ \citenamefont
  {Chan}}]{Begu_i__2024}%
  \BibitemOpen
  \bibfield  {author} {\bibinfo {author} {\bibfnamefont {T.}~\bibnamefont
  {Begušić}}, \bibinfo {author} {\bibfnamefont {J.}~\bibnamefont {Gray}},\
  and\ \bibinfo {author} {\bibfnamefont {G.~K.-L.}\ \bibnamefont {Chan}},\
  }\bibfield  {title} {\bibinfo {title} {Fast and converged classical
  simulations of evidence for the utility of quantum computing before fault
  tolerance},\ }\bibfield  {journal} {\bibinfo  {journal} {Science Advances}\
  }\textbf {\bibinfo {volume} {10}},\ \href
  {https://doi.org/10.1126/sciadv.adk4321} {10.1126/sciadv.adk4321} (\bibinfo
  {year} {2024})\BibitemShut {NoStop}%
\bibitem [{\citenamefont {Verstraete}\ \emph {et~al.}(2008)\citenamefont
  {Verstraete}, \citenamefont {Murg},\ and\ \citenamefont
  {Cirac}}]{Verstraete_2008}%
  \BibitemOpen
  \bibfield  {author} {\bibinfo {author} {\bibfnamefont {F.}~\bibnamefont
  {Verstraete}}, \bibinfo {author} {\bibfnamefont {V.}~\bibnamefont {Murg}},\
  and\ \bibinfo {author} {\bibfnamefont {J.}~\bibnamefont {Cirac}},\ }\bibfield
   {title} {\bibinfo {title} {Matrix product states, projected entangled pair
  states, and variational renormalization group methods for quantum spin
  systems},\ }\href {https://doi.org/10.1080/14789940801912366} {\bibfield
  {journal} {\bibinfo  {journal} {Advances in Physics}\ }\textbf {\bibinfo
  {volume} {57}},\ \bibinfo {pages} {143–224} (\bibinfo {year}
  {2008})}\BibitemShut {NoStop}%
\bibitem [{\citenamefont {Schollwöck}(2011)}]{Schollw_ck_2011}%
  \BibitemOpen
  \bibfield  {author} {\bibinfo {author} {\bibfnamefont {U.}~\bibnamefont
  {Schollwöck}},\ }\bibfield  {title} {\bibinfo {title} {The density-matrix
  renormalization group in the age of matrix product states},\ }\href
  {https://doi.org/10.1016/j.aop.2010.09.012} {\bibfield  {journal} {\bibinfo
  {journal} {Annals of Physics}\ }\textbf {\bibinfo {volume} {326}},\ \bibinfo
  {pages} {96–192} (\bibinfo {year} {2011})}\BibitemShut {NoStop}%
\bibitem [{\citenamefont {Stoudenmire}\ and\ \citenamefont
  {Schwab}(2017)}]{stoudenmire2017}%
  \BibitemOpen
  \bibfield  {author} {\bibinfo {author} {\bibfnamefont {E.~M.}\ \bibnamefont
  {Stoudenmire}}\ and\ \bibinfo {author} {\bibfnamefont {D.~J.}\ \bibnamefont
  {Schwab}},\ }\href {https://arxiv.org/abs/1605.05775} {\bibinfo {title}
  {Supervised learning with quantum-inspired tensor networks}} (\bibinfo {year}
  {2017}),\ \Eprint {https://arxiv.org/abs/1605.05775} {arXiv:1605.05775
  [stat.ML]} \BibitemShut {NoStop}%
\bibitem [{\citenamefont {Wang}\ \emph {et~al.}(2020)\citenamefont {Wang},
  \citenamefont {Roberts}, \citenamefont {Vidal},\ and\ \citenamefont
  {Leichenauer}}]{TNAD_paper}%
  \BibitemOpen
  \bibfield  {author} {\bibinfo {author} {\bibfnamefont {J.}~\bibnamefont
  {Wang}}, \bibinfo {author} {\bibfnamefont {C.}~\bibnamefont {Roberts}},
  \bibinfo {author} {\bibfnamefont {G.}~\bibnamefont {Vidal}},\ and\ \bibinfo
  {author} {\bibfnamefont {S.}~\bibnamefont {Leichenauer}},\ }\href
  {https://arxiv.org/abs/2006.02516} {\bibinfo {title} {Anomaly detection with
  tensor networks}} (\bibinfo {year} {2020}),\ \Eprint
  {https://arxiv.org/abs/2006.02516} {arXiv:2006.02516 [cs.LG]} \BibitemShut
  {NoStop}%
\bibitem [{\citenamefont
  {Evenbly}(2022)}]{evenbly2022practicalguidenumericalimplementation}%
  \BibitemOpen
  \bibfield  {author} {\bibinfo {author} {\bibfnamefont {G.}~\bibnamefont
  {Evenbly}},\ }\href {https://arxiv.org/abs/2202.02138} {\bibinfo {title} {A
  practical guide to the numerical implementation of tensor networks i:
  Contractions, decompositions and gauge freedom}} (\bibinfo {year} {2022}),\
  \Eprint {https://arxiv.org/abs/2202.02138} {arXiv:2202.02138 [quant-ph]}
  \BibitemShut {NoStop}%
\bibitem [{\citenamefont {Zhao}\ \emph {et~al.}(2021)\citenamefont {Zhao},
  \citenamefont {Li}, \citenamefont {Jiang}, \citenamefont {Li}, \citenamefont
  {Li}, \citenamefont {Wang}, \citenamefont {Gong}, \citenamefont {Zhang},\
  and\ \citenamefont {Wei}}]{PhysRevA.104.032603}%
  \BibitemOpen
  \bibfield  {author} {\bibinfo {author} {\bibfnamefont {Y.-Q.}\ \bibnamefont
  {Zhao}}, \bibinfo {author} {\bibfnamefont {R.-G.}\ \bibnamefont {Li}},
  \bibinfo {author} {\bibfnamefont {J.-Z.}\ \bibnamefont {Jiang}}, \bibinfo
  {author} {\bibfnamefont {C.}~\bibnamefont {Li}}, \bibinfo {author}
  {\bibfnamefont {H.-Z.}\ \bibnamefont {Li}}, \bibinfo {author} {\bibfnamefont
  {E.-D.}\ \bibnamefont {Wang}}, \bibinfo {author} {\bibfnamefont {W.-F.}\
  \bibnamefont {Gong}}, \bibinfo {author} {\bibfnamefont {X.}~\bibnamefont
  {Zhang}},\ and\ \bibinfo {author} {\bibfnamefont {Z.-Q.}\ \bibnamefont
  {Wei}},\ }\bibfield  {title} {\bibinfo {title} {Simulation of quantum
  computing on classical supercomputers with tensor-network edge cutting},\
  }\href {https://doi.org/10.1103/PhysRevA.104.032603} {\bibfield  {journal}
  {\bibinfo  {journal} {Phys. Rev. A}\ }\textbf {\bibinfo {volume} {104}},\
  \bibinfo {pages} {032603} (\bibinfo {year} {2021})}\BibitemShut {NoStop}%
\bibitem [{\citenamefont {Cervero~Martín}\ \emph {et~al.}(2023)\citenamefont
  {Cervero~Martín}, \citenamefont {Plekhanov},\ and\ \citenamefont
  {Lubasch}}]{Cervero_Mart_n_2023}%
  \BibitemOpen
  \bibfield  {author} {\bibinfo {author} {\bibfnamefont {E.}~\bibnamefont
  {Cervero~Martín}}, \bibinfo {author} {\bibfnamefont {K.}~\bibnamefont
  {Plekhanov}},\ and\ \bibinfo {author} {\bibfnamefont {M.}~\bibnamefont
  {Lubasch}},\ }\bibfield  {title} {\bibinfo {title} {Barren plateaus in
  quantum tensor network optimization},\ }\href
  {https://doi.org/10.22331/q-2023-04-13-974} {\bibfield  {journal} {\bibinfo
  {journal} {Quantum}\ }\textbf {\bibinfo {volume} {7}},\ \bibinfo {pages}
  {974} (\bibinfo {year} {2023})}\BibitemShut {NoStop}%
\bibitem [{\citenamefont {Liu}\ \emph {et~al.}(2022)\citenamefont {Liu},
  \citenamefont {Yu}, \citenamefont {Duan},\ and\ \citenamefont
  {Deng}}]{PhysRevLett.129.270501}%
  \BibitemOpen
  \bibfield  {author} {\bibinfo {author} {\bibfnamefont {Z.}~\bibnamefont
  {Liu}}, \bibinfo {author} {\bibfnamefont {L.-W.}\ \bibnamefont {Yu}},
  \bibinfo {author} {\bibfnamefont {L.-M.}\ \bibnamefont {Duan}},\ and\
  \bibinfo {author} {\bibfnamefont {D.-L.}\ \bibnamefont {Deng}},\ }\bibfield
  {title} {\bibinfo {title} {Presence and absence of barren plateaus in
  tensor-network based machine learning},\ }\href
  {https://doi.org/10.1103/PhysRevLett.129.270501} {\bibfield  {journal}
  {\bibinfo  {journal} {Phys. Rev. Lett.}\ }\textbf {\bibinfo {volume} {129}},\
  \bibinfo {pages} {270501} (\bibinfo {year} {2022})}\BibitemShut {NoStop}%
\bibitem [{\citenamefont {Menczer}\ \emph {et~al.}(2024)\citenamefont
  {Menczer}, \citenamefont {van Damme}, \citenamefont {Rask}, \citenamefont
  {Huntington}, \citenamefont {Hammond}, \citenamefont {Xantheas},
  \citenamefont {Ganahl},\ and\ \citenamefont {Örs
  Legeza}}]{menczer2024parallelimplementationdensitymatrix}%
  \BibitemOpen
  \bibfield  {author} {\bibinfo {author} {\bibfnamefont {A.}~\bibnamefont
  {Menczer}}, \bibinfo {author} {\bibfnamefont {M.}~\bibnamefont {van Damme}},
  \bibinfo {author} {\bibfnamefont {A.}~\bibnamefont {Rask}}, \bibinfo {author}
  {\bibfnamefont {L.}~\bibnamefont {Huntington}}, \bibinfo {author}
  {\bibfnamefont {J.}~\bibnamefont {Hammond}}, \bibinfo {author} {\bibfnamefont
  {S.~S.}\ \bibnamefont {Xantheas}}, \bibinfo {author} {\bibfnamefont
  {M.}~\bibnamefont {Ganahl}},\ and\ \bibinfo {author} {\bibnamefont {Örs
  Legeza}},\ }\href {https://arxiv.org/abs/2407.07411} {\bibinfo {title}
  {Parallel implementation of the density matrix renormalization group method
  achieving a quarter petaflops performance on a single dgx-h100 gpu node}}
  (\bibinfo {year} {2024}),\ \Eprint {https://arxiv.org/abs/2407.07411}
  {arXiv:2407.07411 [physics.chem-ph]} \BibitemShut {NoStop}%
\bibitem [{\citenamefont {Benenti}\ \emph {et~al.}(2018)\citenamefont
  {Benenti}, \citenamefont {Casati}, \citenamefont {Davide Rossini
  (Universita~di Pisa},\ and\ \citenamefont
  {Strini)}}]{entanglement_correlations}%
  \BibitemOpen
  \bibfield  {author} {\bibinfo {author} {\bibfnamefont {G.}~\bibnamefont
  {Benenti}}, \bibinfo {author} {\bibfnamefont {G.}~\bibnamefont {Casati}},
  \bibinfo {author} {\bibfnamefont {I.}~\bibnamefont {Davide Rossini
  (Universita~di Pisa}},\ and\ \bibinfo {author} {\bibfnamefont
  {G.}~\bibnamefont {Strini)}},\ }\bibinfo {title} {Entanglement and
  non-classical correlations},\ in\ \href
  {https://doi.org/10.1142/9789813237230_0007} {\emph {\bibinfo {booktitle}
  {Principles of Quantum Computation and Information}}}\ (\bibinfo  {publisher}
  {World Scientifics},\ \bibinfo {year} {2018})\ Chap.\ \bibinfo {chapter}
  {Chapter 6}, pp.\ \bibinfo {pages} {241--286}\BibitemShut {NoStop}%
\bibitem [{\citenamefont {Horodecki}\ \emph {et~al.}(2009)\citenamefont
  {Horodecki}, \citenamefont {Horodecki}, \citenamefont {Horodecki},\ and\
  \citenamefont {Horodecki}}]{horodecki}%
  \BibitemOpen
  \bibfield  {author} {\bibinfo {author} {\bibfnamefont {R.}~\bibnamefont
  {Horodecki}}, \bibinfo {author} {\bibfnamefont {P.}~\bibnamefont
  {Horodecki}}, \bibinfo {author} {\bibfnamefont {M.}~\bibnamefont
  {Horodecki}},\ and\ \bibinfo {author} {\bibfnamefont {K.}~\bibnamefont
  {Horodecki}},\ }\bibfield  {title} {\bibinfo {title} {Quantum entanglement},\
  }\href {https://doi.org/10.1103/RevModPhys.81.865} {\bibfield  {journal}
  {\bibinfo  {journal} {Rev. Mod. Phys.}\ }\textbf {\bibinfo {volume} {81}},\
  \bibinfo {pages} {865} (\bibinfo {year} {2009})}\BibitemShut {NoStop}%
\bibitem [{\citenamefont {Horodecki}\ \emph {et~al.}(2024)\citenamefont
  {Horodecki}, \citenamefont {Łukasz Rudnicki},\ and\ \citenamefont
  {Życzkowski}}]{horodecki2024multipartiteentanglement}%
  \BibitemOpen
  \bibfield  {author} {\bibinfo {author} {\bibfnamefont {P.}~\bibnamefont
  {Horodecki}}, \bibinfo {author} {\bibnamefont {Łukasz Rudnicki}},\ and\
  \bibinfo {author} {\bibfnamefont {K.}~\bibnamefont {Życzkowski}},\ }\href
  {https://arxiv.org/abs/2409.04566} {\bibinfo {title} {Multipartite
  entanglement}} (\bibinfo {year} {2024}),\ \Eprint
  {https://arxiv.org/abs/2409.04566} {arXiv:2409.04566 [quant-ph]} \BibitemShut
  {NoStop}%
\bibitem [{\citenamefont {Nielsen}\ and\ \citenamefont
  {Chuang}(2010)}]{Nielsen_Chuang_2010}%
  \BibitemOpen
  \bibfield  {author} {\bibinfo {author} {\bibfnamefont {M.~A.}\ \bibnamefont
  {Nielsen}}\ and\ \bibinfo {author} {\bibfnamefont {I.~L.}\ \bibnamefont
  {Chuang}},\ }\bibfield  {title} {\bibinfo {title} {Introduction to quantum
  mechanics},\ }in\ \href@noop {} {\emph {\bibinfo {booktitle} {Quantum
  Computation and Quantum Information: 10th Anniversary Edition}}}\ (\bibinfo
  {publisher} {{Cambridge University Press}},\ \bibinfo {year} {2010})\ pp.\
  \bibinfo {pages} {60--119}\BibitemShut {NoStop}%
\bibitem [{\citenamefont {Orus}(2014)}]{orusPracticalIntroductionTensor2014}%
  \BibitemOpen
  \bibfield  {author} {\bibinfo {author} {\bibfnamefont {R.}~\bibnamefont
  {Orus}},\ }\bibfield  {title} {\bibinfo {title} {A {{Practical Introduction}}
  to {{Tensor Networks}}: {{Matrix Product States}} and {{Projected Entangled
  Pair States}}},\ }\href {https://doi.org/10.1016/j.aop.2014.06.013}
  {\bibfield  {journal} {\bibinfo  {journal} {Annals of Physics}\ }\textbf
  {\bibinfo {volume} {349}},\ \bibinfo {pages} {117} (\bibinfo {year}
  {2014})},\ \Eprint {https://arxiv.org/abs/1306.2164} {1306.2164} \BibitemShut
  {NoStop}%
\bibitem [{\citenamefont {Sharma}\ \emph {et~al.}(2021)\citenamefont {Sharma},
  \citenamefont {Markopoulos}, \citenamefont {Saber}, \citenamefont {Asif},\
  and\ \citenamefont {Prater-Bennette}}]{9607509}%
  \BibitemOpen
  \bibfield  {author} {\bibinfo {author} {\bibfnamefont {M.}~\bibnamefont
  {Sharma}}, \bibinfo {author} {\bibfnamefont {P.~P.}\ \bibnamefont
  {Markopoulos}}, \bibinfo {author} {\bibfnamefont {E.}~\bibnamefont {Saber}},
  \bibinfo {author} {\bibfnamefont {M.~S.}\ \bibnamefont {Asif}},\ and\
  \bibinfo {author} {\bibfnamefont {A.}~\bibnamefont {Prater-Bennette}},\
  }\bibfield  {title} {\bibinfo {title} {Convolutional auto-encoder with
  tensor-train factorization},\ }in\ \href
  {https://doi.org/10.1109/ICCVW54120.2021.00027} {\emph {\bibinfo {booktitle}
  {2021 IEEE/CVF International Conference on Computer Vision Workshops
  (ICCVW)}}}\ (\bibinfo {year} {2021})\ pp.\ \bibinfo {pages}
  {198--206}\BibitemShut {NoStop}%
\bibitem [{\citenamefont {Su}\ \emph {et~al.}(2024)\citenamefont {Su},
  \citenamefont {Zhou}, \citenamefont {Mo},\ and\ \citenamefont
  {Simonsen}}]{su2024languagemodelingusingtensor}%
  \BibitemOpen
  \bibfield  {author} {\bibinfo {author} {\bibfnamefont {Z.}~\bibnamefont
  {Su}}, \bibinfo {author} {\bibfnamefont {Y.}~\bibnamefont {Zhou}}, \bibinfo
  {author} {\bibfnamefont {F.}~\bibnamefont {Mo}},\ and\ \bibinfo {author}
  {\bibfnamefont {J.~G.}\ \bibnamefont {Simonsen}},\ }\href
  {https://arxiv.org/abs/2405.04590} {\bibinfo {title} {Language modeling using
  tensor trains}} (\bibinfo {year} {2024}),\ \Eprint
  {https://arxiv.org/abs/2405.04590} {arXiv:2405.04590 [cs.CL]} \BibitemShut
  {NoStop}%
\bibitem [{\citenamefont {Oseledets}(2011)}]{tensortrain}%
  \BibitemOpen
  \bibfield  {author} {\bibinfo {author} {\bibfnamefont {I.~V.}\ \bibnamefont
  {Oseledets}},\ }\bibfield  {title} {\bibinfo {title} {Tensor-train
  decomposition},\ }\href {https://doi.org/10.1137/090752286} {\bibfield
  {journal} {\bibinfo  {journal} {SIAM Journal on Scientific Computing}\
  }\textbf {\bibinfo {volume} {33}},\ \bibinfo {pages} {2295} (\bibinfo {year}
  {2011})},\ \Eprint {https://arxiv.org/abs/https://doi.org/10.1137/090752286}
  {https://doi.org/10.1137/090752286} \BibitemShut {NoStop}%
\bibitem [{\citenamefont {Orús}(2019)}]{Orus_2019}%
  \BibitemOpen
  \bibfield  {author} {\bibinfo {author} {\bibfnamefont {R.}~\bibnamefont
  {Orús}},\ }\bibfield  {title} {\bibinfo {title} {Tensor networks for complex
  quantum systems},\ }\href {https://doi.org/10.1038/s42254-019-0086-7}
  {\bibfield  {journal} {\bibinfo  {journal} {Nature Reviews Physics}\ }\textbf
  {\bibinfo {volume} {1}},\ \bibinfo {pages} {538–550} (\bibinfo {year}
  {2019})}\BibitemShut {NoStop}%
\bibitem [{\citenamefont {Affleck}\ \emph {et~al.}(1987)\citenamefont
  {Affleck}, \citenamefont {Kennedy}, \citenamefont {Lieb},\ and\ \citenamefont
  {Tasaki}}]{affleckRigorousResultsValencebond1987}%
  \BibitemOpen
  \bibfield  {author} {\bibinfo {author} {\bibfnamefont {I.}~\bibnamefont
  {Affleck}}, \bibinfo {author} {\bibfnamefont {T.}~\bibnamefont {Kennedy}},
  \bibinfo {author} {\bibfnamefont {E.~H.}\ \bibnamefont {Lieb}},\ and\
  \bibinfo {author} {\bibfnamefont {H.}~\bibnamefont {Tasaki}},\ }\bibfield
  {title} {\bibinfo {title} {Rigorous results on valence-bond ground states in
  antiferromagnets},\ }\href {https://doi.org/10.1103/PhysRevLett.59.799}
  {\bibfield  {journal} {\bibinfo  {journal} {Physical Review Letters}\
  }\textbf {\bibinfo {volume} {59}},\ \bibinfo {pages} {799} (\bibinfo {year}
  {1987})}\BibitemShut {NoStop}%
\bibitem [{\citenamefont {Evenbly}\ and\ \citenamefont
  {Vidal}(2011)}]{evenblyTensorNetworkStates2011}%
  \BibitemOpen
  \bibfield  {author} {\bibinfo {author} {\bibfnamefont {G.}~\bibnamefont
  {Evenbly}}\ and\ \bibinfo {author} {\bibfnamefont {G.}~\bibnamefont
  {Vidal}},\ }\bibfield  {title} {\bibinfo {title} {Tensor {{Network States}}
  and {{Geometry}}},\ }\href {https://doi.org/10.1007/s10955-011-0237-4}
  {\bibfield  {journal} {\bibinfo  {journal} {Journal of Statistical Physics}\
  }\textbf {\bibinfo {volume} {145}},\ \bibinfo {pages} {891} (\bibinfo {year}
  {2011})}\BibitemShut {NoStop}%
\bibitem [{\citenamefont {Eisert}\ \emph {et~al.}(2010)\citenamefont {Eisert},
  \citenamefont {Cramer},\ and\ \citenamefont
  {Plenio}}]{eisertAreaLawsEntanglement2010}%
  \BibitemOpen
  \bibfield  {author} {\bibinfo {author} {\bibfnamefont {J.}~\bibnamefont
  {Eisert}}, \bibinfo {author} {\bibfnamefont {M.}~\bibnamefont {Cramer}},\
  and\ \bibinfo {author} {\bibfnamefont {M.~B.}\ \bibnamefont {Plenio}},\
  }\bibfield  {title} {\bibinfo {title} {Area laws for the entanglement entropy
  - a review},\ }\href {https://doi.org/10.1103/RevModPhys.82.277} {\bibfield
  {journal} {\bibinfo  {journal} {Reviews of Modern Physics}\ }\textbf
  {\bibinfo {volume} {82}},\ \bibinfo {pages} {277} (\bibinfo {year} {2010})},\
  \Eprint {https://arxiv.org/abs/0808.3773} {0808.3773} \BibitemShut {NoStop}%
\bibitem [{\citenamefont {Shi}\ \emph {et~al.}(2006)\citenamefont {Shi},
  \citenamefont {Duan},\ and\ \citenamefont {Vidal}}]{PhysRevA.74.022320}%
  \BibitemOpen
  \bibfield  {author} {\bibinfo {author} {\bibfnamefont {Y.-Y.}\ \bibnamefont
  {Shi}}, \bibinfo {author} {\bibfnamefont {L.-M.}\ \bibnamefont {Duan}},\ and\
  \bibinfo {author} {\bibfnamefont {G.}~\bibnamefont {Vidal}},\ }\bibfield
  {title} {\bibinfo {title} {Classical simulation of quantum many-body systems
  with a tree tensor network},\ }\href
  {https://doi.org/10.1103/PhysRevA.74.022320} {\bibfield  {journal} {\bibinfo
  {journal} {Phys. Rev. A}\ }\textbf {\bibinfo {volume} {74}},\ \bibinfo
  {pages} {022320} (\bibinfo {year} {2006})}\BibitemShut {NoStop}%
\bibitem [{\citenamefont {Vidal}(2007{\natexlab{b}})}]{PhysRevLett.99.220405}%
  \BibitemOpen
  \bibfield  {author} {\bibinfo {author} {\bibfnamefont {G.}~\bibnamefont
  {Vidal}},\ }\bibfield  {title} {\bibinfo {title} {Entanglement
  renormalization},\ }\href {https://doi.org/10.1103/PhysRevLett.99.220405}
  {\bibfield  {journal} {\bibinfo  {journal} {Phys. Rev. Lett.}\ }\textbf
  {\bibinfo {volume} {99}},\ \bibinfo {pages} {220405} (\bibinfo {year}
  {2007}{\natexlab{b}})}\BibitemShut {NoStop}%
\bibitem [{\citenamefont {Okunishi}\ \emph {et~al.}(2023)\citenamefont
  {Okunishi}, \citenamefont {Ueda},\ and\ \citenamefont
  {Nishino}}]{10.1093/ptep/ptad018}%
  \BibitemOpen
  \bibfield  {author} {\bibinfo {author} {\bibfnamefont {K.}~\bibnamefont
  {Okunishi}}, \bibinfo {author} {\bibfnamefont {H.}~\bibnamefont {Ueda}},\
  and\ \bibinfo {author} {\bibfnamefont {T.}~\bibnamefont {Nishino}},\
  }\bibfield  {title} {\bibinfo {title} {Entanglement bipartitioning and tree
  tensor networks},\ }\href {https://doi.org/10.1093/ptep/ptad018} {\bibfield
  {journal} {\bibinfo  {journal} {Progress of Theoretical and Experimental
  Physics}\ }\textbf {\bibinfo {volume} {2023}},\ \bibinfo {pages} {023A02}
  (\bibinfo {year} {2023})},\ \Eprint
  {https://arxiv.org/abs/https://academic.oup.com/ptep/article-pdf/2023/2/023A02/49294184/ptad018.pdf}
  {https://academic.oup.com/ptep/article-pdf/2023/2/023A02/49294184/ptad018.pdf}
  \BibitemShut {NoStop}%
\bibitem [{\citenamefont {Haghshenas}\ \emph {et~al.}(2019)\citenamefont
  {Haghshenas}, \citenamefont {O'Rourke},\ and\ \citenamefont
  {Chan}}]{haghshenas2019conversion}%
  \BibitemOpen
  \bibfield  {author} {\bibinfo {author} {\bibfnamefont {R.}~\bibnamefont
  {Haghshenas}}, \bibinfo {author} {\bibfnamefont {M.~J.}\ \bibnamefont
  {O'Rourke}},\ and\ \bibinfo {author} {\bibfnamefont {G.~K.-L.}\ \bibnamefont
  {Chan}},\ }\bibfield  {title} {\bibinfo {title} {Conversion of projected
  entangled pair states into a canonical form},\ }\href@noop {} {\bibfield
  {journal} {\bibinfo  {journal} {Physical Review B}\ }\textbf {\bibinfo
  {volume} {100}},\ \bibinfo {pages} {054404} (\bibinfo {year}
  {2019})}\BibitemShut {NoStop}%
\bibitem [{\citenamefont {Hyatt}\ and\ \citenamefont
  {Stoudenmire}(2020)}]{hyatt2020dmrgapproachoptimizingtwodimensional}%
  \BibitemOpen
  \bibfield  {author} {\bibinfo {author} {\bibfnamefont {K.}~\bibnamefont
  {Hyatt}}\ and\ \bibinfo {author} {\bibfnamefont {E.~M.}\ \bibnamefont
  {Stoudenmire}},\ }\href {https://arxiv.org/abs/1908.08833} {\bibinfo {title}
  {Dmrg approach to optimizing two-dimensional tensor networks}} (\bibinfo
  {year} {2020}),\ \Eprint {https://arxiv.org/abs/1908.08833} {arXiv:1908.08833
  [cond-mat.str-el]} \BibitemShut {NoStop}%
\bibitem [{\citenamefont {Hikihara}\ \emph {et~al.}(2023)\citenamefont
  {Hikihara}, \citenamefont {Ueda}, \citenamefont {Okunishi}, \citenamefont
  {Harada},\ and\ \citenamefont {Nishino}}]{PhysRevResearch.5.013031}%
  \BibitemOpen
  \bibfield  {author} {\bibinfo {author} {\bibfnamefont {T.}~\bibnamefont
  {Hikihara}}, \bibinfo {author} {\bibfnamefont {H.}~\bibnamefont {Ueda}},
  \bibinfo {author} {\bibfnamefont {K.}~\bibnamefont {Okunishi}}, \bibinfo
  {author} {\bibfnamefont {K.}~\bibnamefont {Harada}},\ and\ \bibinfo {author}
  {\bibfnamefont {T.}~\bibnamefont {Nishino}},\ }\bibfield  {title} {\bibinfo
  {title} {Automatic structural optimization of tree tensor networks},\ }\href
  {https://doi.org/10.1103/PhysRevResearch.5.013031} {\bibfield  {journal}
  {\bibinfo  {journal} {Phys. Rev. Res.}\ }\textbf {\bibinfo {volume} {5}},\
  \bibinfo {pages} {013031} (\bibinfo {year} {2023})}\BibitemShut {NoStop}%
\bibitem [{\citenamefont {Hikihara}\ \emph {et~al.}(2024)\citenamefont
  {Hikihara}, \citenamefont {Ueda}, \citenamefont {Okunishi}, \citenamefont
  {Harada},\ and\ \citenamefont
  {Nishino}}]{hikihara2024visualizationentanglementgeometrystructural}%
  \BibitemOpen
  \bibfield  {author} {\bibinfo {author} {\bibfnamefont {T.}~\bibnamefont
  {Hikihara}}, \bibinfo {author} {\bibfnamefont {H.}~\bibnamefont {Ueda}},
  \bibinfo {author} {\bibfnamefont {K.}~\bibnamefont {Okunishi}}, \bibinfo
  {author} {\bibfnamefont {K.}~\bibnamefont {Harada}},\ and\ \bibinfo {author}
  {\bibfnamefont {T.}~\bibnamefont {Nishino}},\ }\href
  {https://arxiv.org/abs/2401.16000} {\bibinfo {title} {Visualization of
  entanglement geometry by structural optimization of tree tensor network}}
  (\bibinfo {year} {2024}),\ \Eprint {https://arxiv.org/abs/2401.16000}
  {arXiv:2401.16000 [cond-mat.stat-mech]} \BibitemShut {NoStop}%
\bibitem [{\citenamefont {Lu}\ \emph {et~al.}(2023)\citenamefont {Lu},
  \citenamefont {Gong}, \citenamefont {Ye}, \citenamefont {Zhang},\ and\
  \citenamefont {Zhang}}]{lu2023surveymachinelearningsamples}%
  \BibitemOpen
  \bibfield  {author} {\bibinfo {author} {\bibfnamefont {J.}~\bibnamefont
  {Lu}}, \bibinfo {author} {\bibfnamefont {P.}~\bibnamefont {Gong}}, \bibinfo
  {author} {\bibfnamefont {J.}~\bibnamefont {Ye}}, \bibinfo {author}
  {\bibfnamefont {J.}~\bibnamefont {Zhang}},\ and\ \bibinfo {author}
  {\bibfnamefont {C.}~\bibnamefont {Zhang}},\ }\href
  {https://arxiv.org/abs/2009.02653} {\bibinfo {title} {A survey on machine
  learning from few samples}} (\bibinfo {year} {2023}),\ \Eprint
  {https://arxiv.org/abs/2009.02653} {arXiv:2009.02653 [cs.LG]} \BibitemShut
  {NoStop}%
\bibitem [{\citenamefont {Fuksa}\ \emph {et~al.}(2024)\citenamefont {Fuksa},
  \citenamefont {Götte}, \citenamefont {Roth},\ and\ \citenamefont
  {Eisert}}]{Fuksa_2024}%
  \BibitemOpen
  \bibfield  {author} {\bibinfo {author} {\bibfnamefont {J.}~\bibnamefont
  {Fuksa}}, \bibinfo {author} {\bibfnamefont {M.}~\bibnamefont {Götte}},
  \bibinfo {author} {\bibfnamefont {I.}~\bibnamefont {Roth}},\ and\ \bibinfo
  {author} {\bibfnamefont {J.}~\bibnamefont {Eisert}},\ }\bibfield  {title}
  {\bibinfo {title} {A quantum inspired approach to learning dynamical laws
  from data—block-sparsity and gauge-mediated weight sharing},\ }\href
  {https://doi.org/10.1088/2632-2153/ad4f4e} {\bibfield  {journal} {\bibinfo
  {journal} {Machine Learning: Science and Technology}\ }\textbf {\bibinfo
  {volume} {5}},\ \bibinfo {pages} {025064} (\bibinfo {year}
  {2024})}\BibitemShut {NoStop}%
\bibitem [{\citenamefont {Cerezo}\ \emph {et~al.}(2022)\citenamefont {Cerezo},
  \citenamefont {Verdon}, \citenamefont {Huang}, \citenamefont {Cincio},\ and\
  \citenamefont {Coles}}]{Cerezo_2022}%
  \BibitemOpen
  \bibfield  {author} {\bibinfo {author} {\bibfnamefont {M.}~\bibnamefont
  {Cerezo}}, \bibinfo {author} {\bibfnamefont {G.}~\bibnamefont {Verdon}},
  \bibinfo {author} {\bibfnamefont {H.-Y.}\ \bibnamefont {Huang}}, \bibinfo
  {author} {\bibfnamefont {L.}~\bibnamefont {Cincio}},\ and\ \bibinfo {author}
  {\bibfnamefont {P.~J.}\ \bibnamefont {Coles}},\ }\bibfield  {title} {\bibinfo
  {title} {Challenges and opportunities in quantum machine learning},\ }\href
  {https://doi.org/10.1038/s43588-022-00311-3} {\bibfield  {journal} {\bibinfo
  {journal} {Nature Computational Science}\ }\textbf {\bibinfo {volume} {2}},\
  \bibinfo {pages} {567–576} (\bibinfo {year} {2022})}\BibitemShut {NoStop}%
\bibitem [{\citenamefont {Wu}\ \emph {et~al.}(2018)\citenamefont {Wu},
  \citenamefont {Di}, \citenamefont {Cappello}, \citenamefont {Finkel},
  \citenamefont {Alexeev},\ and\ \citenamefont
  {Chong}}]{wu2018memoryefficientquantumcircuitsimulation}%
  \BibitemOpen
  \bibfield  {author} {\bibinfo {author} {\bibfnamefont {X.-C.}\ \bibnamefont
  {Wu}}, \bibinfo {author} {\bibfnamefont {S.}~\bibnamefont {Di}}, \bibinfo
  {author} {\bibfnamefont {F.}~\bibnamefont {Cappello}}, \bibinfo {author}
  {\bibfnamefont {H.}~\bibnamefont {Finkel}}, \bibinfo {author} {\bibfnamefont
  {Y.}~\bibnamefont {Alexeev}},\ and\ \bibinfo {author} {\bibfnamefont {F.~T.}\
  \bibnamefont {Chong}},\ }\href {https://arxiv.org/abs/1811.05630} {\bibinfo
  {title} {Memory-efficient quantum circuit simulation by using lossy data
  compression}} (\bibinfo {year} {2018}),\ \Eprint
  {https://arxiv.org/abs/1811.05630} {arXiv:1811.05630 [quant-ph]} \BibitemShut
  {NoStop}%
\bibitem [{\citenamefont {Pan}\ \emph {et~al.}(2024)\citenamefont {Pan},
  \citenamefont {Gu}, \citenamefont {Kuang}, \citenamefont {Liu},\ and\
  \citenamefont {Zhang}}]{pan2024efficientquantumcircuitsimulation}%
  \BibitemOpen
  \bibfield  {author} {\bibinfo {author} {\bibfnamefont {F.}~\bibnamefont
  {Pan}}, \bibinfo {author} {\bibfnamefont {H.}~\bibnamefont {Gu}}, \bibinfo
  {author} {\bibfnamefont {L.}~\bibnamefont {Kuang}}, \bibinfo {author}
  {\bibfnamefont {B.}~\bibnamefont {Liu}},\ and\ \bibinfo {author}
  {\bibfnamefont {P.}~\bibnamefont {Zhang}},\ }\href
  {https://arxiv.org/abs/2310.03978} {\bibinfo {title} {Efficient quantum
  circuit simulation by tensor network methods on modern gpus}} (\bibinfo
  {year} {2024}),\ \Eprint {https://arxiv.org/abs/2310.03978} {arXiv:2310.03978
  [quant-ph]} \BibitemShut {NoStop}%
\bibitem [{\citenamefont {Sanchez-Ramirez}\ \emph {et~al.}(2021)\citenamefont
  {Sanchez-Ramirez}, \citenamefont {Conejero}, \citenamefont {Lordan},
  \citenamefont {Queralt}, \citenamefont {Cortes}, \citenamefont {Badia},\ and\
  \citenamefont {Garcia-Saez}}]{Sanchez_Ramirez_2021}%
  \BibitemOpen
  \bibfield  {author} {\bibinfo {author} {\bibfnamefont {S.}~\bibnamefont
  {Sanchez-Ramirez}}, \bibinfo {author} {\bibfnamefont {J.}~\bibnamefont
  {Conejero}}, \bibinfo {author} {\bibfnamefont {F.}~\bibnamefont {Lordan}},
  \bibinfo {author} {\bibfnamefont {A.}~\bibnamefont {Queralt}}, \bibinfo
  {author} {\bibfnamefont {T.}~\bibnamefont {Cortes}}, \bibinfo {author}
  {\bibfnamefont {R.~M.}\ \bibnamefont {Badia}},\ and\ \bibinfo {author}
  {\bibfnamefont {A.}~\bibnamefont {Garcia-Saez}},\ }\bibfield  {title}
  {\bibinfo {title} {Rosnet: A block tensor algebra library for out-of-core
  quantum computing simulation},\ }in\ \href
  {https://doi.org/10.1109/qcs54837.2021.00004} {\emph {\bibinfo {booktitle}
  {2021 IEEE/ACM Second International Workshop on Quantum Computing Software
  (QCS)}}}\ (\bibinfo  {publisher} {IEEE},\ \bibinfo {year} {2021})\ p.\
  \bibinfo {pages} {1–8}\BibitemShut {NoStop}%
\bibitem [{\citenamefont {Vidal}(2004)}]{PhysRevLett.93.040502}%
  \BibitemOpen
  \bibfield  {author} {\bibinfo {author} {\bibfnamefont {G.}~\bibnamefont
  {Vidal}},\ }\bibfield  {title} {\bibinfo {title} {Efficient simulation of
  one-dimensional quantum many-body systems},\ }\href
  {https://doi.org/10.1103/PhysRevLett.93.040502} {\bibfield  {journal}
  {\bibinfo  {journal} {Phys. Rev. Lett.}\ }\textbf {\bibinfo {volume} {93}},\
  \bibinfo {pages} {040502} (\bibinfo {year} {2004})}\BibitemShut {NoStop}%
\bibitem [{\citenamefont {Hashizume}\ \emph {et~al.}(2020)\citenamefont
  {Hashizume}, \citenamefont {Halimeh},\ and\ \citenamefont
  {McCulloch}}]{PhysRevB.102.035115}%
  \BibitemOpen
  \bibfield  {author} {\bibinfo {author} {\bibfnamefont {T.}~\bibnamefont
  {Hashizume}}, \bibinfo {author} {\bibfnamefont {J.~C.}\ \bibnamefont
  {Halimeh}},\ and\ \bibinfo {author} {\bibfnamefont {I.~P.}\ \bibnamefont
  {McCulloch}},\ }\bibfield  {title} {\bibinfo {title} {Hybrid infinite
  time-evolving block decimation algorithm for long-range multidimensional
  quantum many-body systems},\ }\href
  {https://doi.org/10.1103/PhysRevB.102.035115} {\bibfield  {journal} {\bibinfo
   {journal} {Phys. Rev. B}\ }\textbf {\bibinfo {volume} {102}},\ \bibinfo
  {pages} {035115} (\bibinfo {year} {2020})}\BibitemShut {NoStop}%
\bibitem [{\citenamefont {McClean}\ \emph {et~al.}(2018)\citenamefont
  {McClean}, \citenamefont {Boixo}, \citenamefont {Smelyanskiy}, \citenamefont
  {Babbush},\ and\ \citenamefont {Neven}}]{McClean_2018}%
  \BibitemOpen
  \bibfield  {author} {\bibinfo {author} {\bibfnamefont {J.~R.}\ \bibnamefont
  {McClean}}, \bibinfo {author} {\bibfnamefont {S.}~\bibnamefont {Boixo}},
  \bibinfo {author} {\bibfnamefont {V.~N.}\ \bibnamefont {Smelyanskiy}},
  \bibinfo {author} {\bibfnamefont {R.}~\bibnamefont {Babbush}},\ and\ \bibinfo
  {author} {\bibfnamefont {H.}~\bibnamefont {Neven}},\ }\bibfield  {title}
  {\bibinfo {title} {Barren plateaus in quantum neural network training
  landscapes},\ }\bibfield  {journal} {\bibinfo  {journal} {Nature
  Communications}\ }\textbf {\bibinfo {volume} {9}},\ \href
  {https://doi.org/10.1038/s41467-018-07090-4} {10.1038/s41467-018-07090-4}
  (\bibinfo {year} {2018})\BibitemShut {NoStop}%
\bibitem [{\citenamefont {Cerezo}\ \emph {et~al.}(2024)\citenamefont {Cerezo},
  \citenamefont {Larocca}, \citenamefont {García-Martín}, \citenamefont
  {Diaz}, \citenamefont {Braccia}, \citenamefont {Fontana}, \citenamefont
  {Rudolph}, \citenamefont {Bermejo}, \citenamefont {Ijaz}, \citenamefont
  {Thanasilp}, \citenamefont {Anschuetz},\ and\ \citenamefont
  {Holmes}}]{cerezo2024doesprovableabsencebarren}%
  \BibitemOpen
  \bibfield  {author} {\bibinfo {author} {\bibfnamefont {M.}~\bibnamefont
  {Cerezo}}, \bibinfo {author} {\bibfnamefont {M.}~\bibnamefont {Larocca}},
  \bibinfo {author} {\bibfnamefont {D.}~\bibnamefont {García-Martín}},
  \bibinfo {author} {\bibfnamefont {N.~L.}\ \bibnamefont {Diaz}}, \bibinfo
  {author} {\bibfnamefont {P.}~\bibnamefont {Braccia}}, \bibinfo {author}
  {\bibfnamefont {E.}~\bibnamefont {Fontana}}, \bibinfo {author} {\bibfnamefont
  {M.~S.}\ \bibnamefont {Rudolph}}, \bibinfo {author} {\bibfnamefont
  {P.}~\bibnamefont {Bermejo}}, \bibinfo {author} {\bibfnamefont
  {A.}~\bibnamefont {Ijaz}}, \bibinfo {author} {\bibfnamefont {S.}~\bibnamefont
  {Thanasilp}}, \bibinfo {author} {\bibfnamefont {E.~R.}\ \bibnamefont
  {Anschuetz}},\ and\ \bibinfo {author} {\bibfnamefont {Z.}~\bibnamefont
  {Holmes}},\ }\href {https://arxiv.org/abs/2312.09121} {\bibinfo {title} {Does
  provable absence of barren plateaus imply classical simulability? or, why we
  need to rethink variational quantum computing}} (\bibinfo {year} {2024}),\
  \Eprint {https://arxiv.org/abs/2312.09121} {arXiv:2312.09121 [quant-ph]}
  \BibitemShut {NoStop}%
\bibitem [{\citenamefont {Schuld}\ and\ \citenamefont
  {Killoran}(2022)}]{Schuld_2022}%
  \BibitemOpen
  \bibfield  {author} {\bibinfo {author} {\bibfnamefont {M.}~\bibnamefont
  {Schuld}}\ and\ \bibinfo {author} {\bibfnamefont {N.}~\bibnamefont
  {Killoran}},\ }\bibfield  {title} {\bibinfo {title} {Is quantum advantage the
  right goal for quantum machine learning?},\ }\bibfield  {journal} {\bibinfo
  {journal} {PRX Quantum}\ }\textbf {\bibinfo {volume} {3}},\ \href
  {https://doi.org/10.1103/prxquantum.3.030101} {10.1103/prxquantum.3.030101}
  (\bibinfo {year} {2022})\BibitemShut {NoStop}%
\bibitem [{\citenamefont {Basheer}\ \emph {et~al.}(2024)\citenamefont
  {Basheer}, \citenamefont {Feng}, \citenamefont {Ferrie}, \citenamefont {Li},\
  and\ \citenamefont
  {Pashayan}}]{basheer2024trainabilityclassicalsimulabilitylearning}%
  \BibitemOpen
  \bibfield  {author} {\bibinfo {author} {\bibfnamefont {A.}~\bibnamefont
  {Basheer}}, \bibinfo {author} {\bibfnamefont {Y.}~\bibnamefont {Feng}},
  \bibinfo {author} {\bibfnamefont {C.}~\bibnamefont {Ferrie}}, \bibinfo
  {author} {\bibfnamefont {S.}~\bibnamefont {Li}},\ and\ \bibinfo {author}
  {\bibfnamefont {H.}~\bibnamefont {Pashayan}},\ }\href
  {https://arxiv.org/abs/2409.10055} {\bibinfo {title} {On the trainability and
  classical simulability of learning matrix product states variationally}}
  (\bibinfo {year} {2024}),\ \Eprint {https://arxiv.org/abs/2409.10055}
  {arXiv:2409.10055 [quant-ph]} \BibitemShut {NoStop}%
\bibitem [{\citenamefont {Baydin}\ \emph {et~al.}(2018)\citenamefont {Baydin},
  \citenamefont {Pearlmutter}, \citenamefont {Radul},\ and\ \citenamefont
  {Siskind}}]{baydin2018automaticdifferentiationmachinelearning}%
  \BibitemOpen
  \bibfield  {author} {\bibinfo {author} {\bibfnamefont {A.~G.}\ \bibnamefont
  {Baydin}}, \bibinfo {author} {\bibfnamefont {B.~A.}\ \bibnamefont
  {Pearlmutter}}, \bibinfo {author} {\bibfnamefont {A.~A.}\ \bibnamefont
  {Radul}},\ and\ \bibinfo {author} {\bibfnamefont {J.~M.}\ \bibnamefont
  {Siskind}},\ }\href {https://arxiv.org/abs/1502.05767} {\bibinfo {title}
  {Automatic differentiation in machine learning: a survey}} (\bibinfo {year}
  {2018}),\ \Eprint {https://arxiv.org/abs/1502.05767} {arXiv:1502.05767
  [cs.SC]} \BibitemShut {NoStop}%
\bibitem [{\citenamefont {Gray}(2018)}]{Gray2018}%
  \BibitemOpen
  \bibfield  {author} {\bibinfo {author} {\bibfnamefont {J.}~\bibnamefont
  {Gray}},\ }\bibfield  {title} {\bibinfo {title} {quimb: A python package for
  quantum information and many-body calculations},\ }\href
  {https://doi.org/10.21105/joss.00819} {\bibfield  {journal} {\bibinfo
  {journal} {Journal of Open Source Software}\ }\textbf {\bibinfo {volume}
  {3}},\ \bibinfo {pages} {819} (\bibinfo {year} {2018})}\BibitemShut {NoStop}%
\bibitem [{\citenamefont {Bradbury}\ \emph {et~al.}(2018)\citenamefont
  {Bradbury}, \citenamefont {Frostig}, \citenamefont {Hawkins}, \citenamefont
  {Johnson}, \citenamefont {Leary}, \citenamefont {Maclaurin}, \citenamefont
  {Necula}, \citenamefont {Paszke}, \citenamefont {Vander{P}las}, \citenamefont
  {Wanderman-{M}ilne},\ and\ \citenamefont {Zhang}}]{jax2018github}%
  \BibitemOpen
  \bibfield  {author} {\bibinfo {author} {\bibfnamefont {J.}~\bibnamefont
  {Bradbury}}, \bibinfo {author} {\bibfnamefont {R.}~\bibnamefont {Frostig}},
  \bibinfo {author} {\bibfnamefont {P.}~\bibnamefont {Hawkins}}, \bibinfo
  {author} {\bibfnamefont {M.~J.}\ \bibnamefont {Johnson}}, \bibinfo {author}
  {\bibfnamefont {C.}~\bibnamefont {Leary}}, \bibinfo {author} {\bibfnamefont
  {D.}~\bibnamefont {Maclaurin}}, \bibinfo {author} {\bibfnamefont
  {G.}~\bibnamefont {Necula}}, \bibinfo {author} {\bibfnamefont
  {A.}~\bibnamefont {Paszke}}, \bibinfo {author} {\bibfnamefont
  {J.}~\bibnamefont {Vander{P}las}}, \bibinfo {author} {\bibfnamefont
  {S.}~\bibnamefont {Wanderman-{M}ilne}},\ and\ \bibinfo {author}
  {\bibfnamefont {Q.}~\bibnamefont {Zhang}},\ }\href
  {http://github.com/jax-ml/jax} {\bibinfo {title} {{JAX}: composable
  transformations of {P}ython+{N}um{P}y programs}} (\bibinfo {year}
  {2018})\BibitemShut {NoStop}%
\bibitem [{https://www.bsc.es/marenostrum/marenostrum-5()}]{mn5}%
  \BibitemOpen
  https://www.bsc.es/marenostrum/marenostrum-5,\ \href@noop {} {}\BibitemShut
  {NoStop}%
\end{thebibliography}%





\end{document}